\documentclass[twocolumn,showpacs,preprintnumbers,amsmath,amssymb]{revtex4} 

\usepackage{graphicx}
\usepackage{dcolumn}
\usepackage{bm}

\begin{document}

\title{Algebraic Approach to Bare Nucleon Matrix Elements of Quark 
Operators 
\footnote{Dedicated to the memory of Professor Gerhard Soff (1949 - 2004).}}

\author{Sven Zschocke}
\affiliation{Forschungszentrum Rossendorf, Postfach 51 01 19, 
01314 Dresden, Germany}
\affiliation{Institut f\"ur Theoretische Physik,
Technische Universit\"at Dresden, 01062 Dresden, Germany}
\affiliation{Section for Theoretical Physics,  
University of Bergen, Allegaten 55, 5007 Bergen, Norway}
\author{Burkhard K\"ampfer}
\affiliation{Forschungszentrum Rossendorf, Postfach 51 01 19, 
01314 Dresden, Germany}
\author{G\"unter Plunien}
\affiliation{Institut f\"ur Theoretische Physik, 
Technische Universit\"at Dresden, 01062 Dresden, Germany}

\begin{abstract}
An algebraic method for evaluating bare nucleon matrix elements of quark
operators is proposed. Thereby, bare nucleon matrix elements are traced back to
vacuum matrix elements. The method is similar to the soft pion theorem.
Matrix elements of two-quark, four-quark and
six-quark operators inside the bare nucleon are considered.
\end{abstract}

\maketitle

\section{\label{sec:chapter1}Introduction}

One ultimate goal of contemporary strong interaction physics is to find a
comprehension of the
physical properties of hadrons by means of the underlying theory of Quantum
Chromodynamics (QCD).
Among several methods which provide a link between QCD (quark and gluon)
degrees of freedom and the hadronic spectrum are the QCD sum rules which have
to be considered as important nonperturbative approach in understanding the
physical observables of hadrons. The sum rule method, first developed
for the vacuum \cite{sumrule}, has later been extended to finite density
\cite{sumrule_n_5,sumrule_n_10, sumrule_n_15}, finite temperature
\cite{sumrule_T_5, sumrule_T_10}, and mixed finite density and 
finite temperature 
\cite{density_temperature}. Within the QCD sum rule approach, and more
generally in hadron physics, pions and nucleons have to be considered as
important degrees of freedom because the pion is the lightest
(Goldstone) meson, while the nucleon is the lightest baryon.
In-medium QCD sum rules provide a direct way to relate changes of
hadronic properties to changes of the various condensates, i.e. nucleon
and pion expectation values of quark and gluon fields.
Therefore, expectation
values of a local operator $\hat{\cal O}$ taken between these states,
$\langle \pi_{\rm phys} | \hat{\cal O} | \pi_{\rm phys}\rangle$ and
$\langle N_{\rm phys} | \hat{\cal O} | N_{\rm phys}\rangle$,
need to be known.
However, the predictive power of the QCD sum rule method in matter meets
uncertainties when evaluating condensates, especially higher mass dimension
condensates inside the nucleon. Accordingly, the exploration
of nucleon matrix elements is presently an active field of
hadron physics, cf. \cite{matrixlement_nucleon_5, lattice_1}.

If the operator $\hat {\cal O}$ consists of hadronic fields, then in principle
one needs an effective hadronic theory which decribes the interaction
between pions and nucleons, respectively, and the hadrons from which the
operator  $\hat {\cal O}$ is made of for evaluating these matrix elements.
However, if one is concerned with pion matrix elements then the use
of soft pion theorems \cite{lit1,lit3, lit4, lit5, hosaka} gives
in general good estimates for such expressions, which are related to several
so called low-energy theorems like Goldberger-Treiman relation
\cite{Goldberger_Treiman}, Adler-Weisberger sum rule \cite{Adler-Weisberger}
or Cabibbo-Radicati sum rule \cite{Cabibbo-Radicati}.
These soft pion theorems as algebraic tools are based on the hypothesis of
partially conserved axial vector current (PCAC) \cite{pcac1, pcac2, pcac3} and
postulated current algebra commutation relations \cite{lit4, lit5},
and allow in general to
trace the pion matrix elements of operators made of effective hadronic fields
back to vacuum matrix elements. A feature of the soft pion theorems is that
they can also be deduced within quark degrees of freedom.
Accordingly, pion matrix elements of quark field
operators have also been evaluated by means of the soft pion theorem
(if we speak about the soft pion theorem then we mean the special theorem
considered in the Appendix~\ref{sec:chapter7} 
which is the relevant one in our context)
expressing the pion field and axial
vector current, respectively, by interpolating fields made of quark degrees
of freedom \cite{Hatsuda, pion}.

After discovering the powerful method of current algebra for mesons
several attempts have been made to investigate the possibilities for extending
this algebra to the case of baryons. Especially, the analog hypothesis of a
partially conserved baryon current (PCBC) and the related (and postulated) 
baryon current algebra has been investigated long time ago
\cite{pcbc1, pcbc2, pcbc3, pcbc4, pcbc5, pcbc6}. These attempts focussed on
the construction of baryon currents by products of nucleon fields. Furthermore,
in \cite{pcbc7} this procedure has been studied by considering
baryon currents made of quark degrees of freedom where several relations
between form factors, e.g. baryon-meson vertex form factors,
have been obtained. However, it turned out that, while the PCAC
directly leads to the mentioned soft pion theorems for evaluating pion matrix
elements, the PCBC does not provide a comprehensive algebraic theorem for
evaluating
nucleon matrix elements. Therefore, up to now for evaluating nucleon matrix
elements of an operator $\hat {\cal O}$ consisting of quark fields
more involved tools are needed
like chiral quark model \cite{matrixlement_nucleon_5}, lattice evaluations
\cite{lattice_1}, or Nambu-Jona-Lasinio model \cite{four_5}.
From this point of view it seems very tempting to look for an algebraic
approach for evaluating nucleon matrix elements in analogy to the soft pion
theorem. Here, by using directly the nucleon field instead the nucleon
current, we propose such an algebraic approach for evaluating matrix
elements of quark operators taken between a bare nucleon, i.e.
the valence quark contribution.

To clarify what the terminology "bare nucleon" means
we recall the basic QCD structure of nucleons.
From deep inelastic lepton-nucleon scattering (DIS) experiments we know that
nucleons are composite color-singlet systems made of partons. In the language
of QCD these are three valence quarks with a current quark mass,
accompanied by virtual sea quarks and gluons.
Accordingly, the physical nucleon state $|N_{\rm phys}\rangle$ is a
highly complicated object consisting of many configurations in the Fock space.
For instance, in the case of the proton, the Fock
expansion begins with the color-singlet state $| u u d\rangle$ consisting
of three valence quarks which is the so called bare proton state,
and continues with $| u u d g\rangle$, $| u u d \overline{q} q \rangle$
and further sea quark and gluon states that span the degrees of freedom of
the proton in QCD.

In the low energy region, many properties of the nucleon can rather
successfully be described by approximating the virtual sea quarks and gluons by
a cloud of mesons, especially pions, surrounding the bare valence quark core.
Accordingly, in the pion cloud model, which resembles the Tamm-Dancoff method
\cite{Tamm_Dancoff1,Tamm_Dancoff2, Tamm_Dankoff1, Tamm_Dankoff2, 
Tamm_Dankoff5, Tamm_Dankoff6} 
the physical nucleon is viewed as a bare nucleon,
which accounts for the three valence quarks,  
accompanied by the pion cloud
which accounts for the virtual sea quarks and gluons.
Then the Fock representation for the physical nucleon reads 
\cite{Tamm_Dankoff1, Tamm_Dankoff2, Tamm_Dankoff3, tamm1, tamm2, Z_Constant}
\begin{eqnarray}
| N_{\rm phys} \rangle = Z_N^{1/2} \left( | N \rangle +
\phi_{1} \; 
| N \pi \rangle + \phi_{2} \; | N \pi \pi \rangle + ... \right) \;,  
\label{bare_nucleon_5}
\end{eqnarray}
where the Fock state $| N \rangle$ represents a bare nucleon state,
$| N \pi \rangle$ and $| N \pi \pi \rangle$ represent a bare nucleon with one
pion and two pions, respectively,
and the dots stand for all of the Fock states consisting of one
bare nucleon with more than one pion or heavier mesons.
The probability amplitudes $\phi_{n}$ to find the nucleon in the state
$| N \, n \pi \rangle$ can be evaluated by using a Hamiltonian which
describes the pion-nucleon interaction \cite{Tamm_Dancoff1, Tamm_Dancoff2,
Tamm_Dankoff1, Tamm_Dankoff2,Tamm_Dankoff3}.
Then the bare nucleon probability can also be determined and turns out to be
$Z_N \simeq 0.9$ \cite{Tamm_Dancoff2, tamm1}.
Since the deviation of $Z_N$ from $1$ comes from pion-nucleon interaction
one has to put $Z_N = 1$ if the pion cloud is not taken into account.
By using the Fock expansion
(\ref{bare_nucleon_5}) the expectation value
of an observable $\hat{\cal O}$ taken between the physical nucleon states
is given by \cite{Tamm_Dancoff2, Tamm_Dankoff1, footnote1}, 
\begin{eqnarray}
\langle N_{\rm phys} | \hat{\cal{O}} | N_{\rm phys} \rangle
= Z_N \bigg( \langle N | \hat{\cal O} | N \rangle +
\nonumber\\
\phi_{1}^2 \; \langle N \pi | \hat{\cal O} | N \pi \rangle
+ \phi_{2}^2 \; \langle N \pi \pi | \hat{\cal O} | N \pi \pi \rangle
+ ...\bigg) \;.
\label{bare_nucleon_10}
\end{eqnarray}
The first term on the right side of (\ref{bare_nucleon_10}), i.e. the
contribution of the bare nucleon without pions, plays an important role
for two reasons.
First, the bare nucleon is expected to give the main contribution in many
cases \cite{footnote2}.
And second, for the leading chiral correction
one needs only the contributions of the lowest-momentum pions in the cloud
allowing an application of the soft pion theorem 
(see Appendix~\ref{sec:chapter7}),
which then reduces the pion cloud terms in (\ref{bare_nucleon_10})
also to bare nucleon matrix elements \cite{krippa1, krippa2}.
Accordingly, in this paper we focus on bare nucleon matrix elements
and propose an algebraic method for evaluating them.
This approach seems capable to estimate nucleon matrix elements of
quark operators in a
straightforward way. We also note that within the algebraic approach
new parameters are not necessary since the bare nucleon
matrix elements are traced back to va\~cuum matrix elements, like in the
soft pion theorem. We apply the method on two-quark, four-quark and, finally, 
on six-quark operators 
inside the nucleon which so far have not been evaluated.

The paper is organized as follows.
In section~\ref{sec:chapter2} 
we derive an algebraic formula for evaluating matrix elements
taken between the state of a bare nucleon.
In section~\ref{sec:chapter3} a valence quark field operator with the
quantum numbers  of a bare nucleon is introduced.
A few tests of the nucleon formula on well known bare nucleon matrix
elements of two-quark operators are given in section~\ref{sec:chapter4A} 
(currents) and
\ref{sec:chapter4B} (chiral condensate).
In section~\ref{sec:chapter4C} we explore the valence quark contribution of
four-quark condensates within the algebraic method developed and assert an
interesting agreement with the results of groundstate saturation approximation
when taking properly the valence quark contribution. 
We also compare our findings 
for the valence quark contribution of four-quark condensates with recently
obtained results within a chiral quark model.
In section~\ref{sec:chapter5} 
we evaluate six-quark condensates inside the bare nucleon.
A summary of the results and an outlook can be found in 
section~\ref{sec:chapter6}.
In Appendix~\ref{sec:chapter7} 
a derivation of the soft pion theorem is given which
shows the similarity of it with our algebraic approach.
Details of some evaluations are relegated to the Appendix~\ref{sec:chapter8}.

\section{\label{sec:chapter2}Nucleon formula}

Let $\hat{{\cal O}} (x)$ be a local operator which may depend on 
space and time, $x = ({\bf r}, t)$. We are interested in matrix elements taken 
between two bare nucleon states $| N (k, \sigma) \rangle$ with four-momentum 
$k$ and spin $\sigma$ (i.e. $| N \rangle$ is either a bare proton
$| p \rangle$ or a bare neutron $| n \rangle$ state, which are considered as
QCD eigenstates). 
To derive a formula for such matrix elements between bare nucleons with finite 
nucleon masses and momenta we first apply the Lehmann-Symanzik-Zimmermann (LSZ) 
reduction \cite{LSZ, Zuber} on one nucleon state,  
\begin{eqnarray}
&&\langle N (k_2, \sigma_2) | \hat{\cal O} (x) | N (k_1, \sigma_1) \rangle
\nonumber\\
&& = i Z_{\Psi}^{-1/2} \int d^4 x_1
\langle N (k_2, \sigma_2) | \;{\rm T}_W \left(\hat{\cal O} (x)
\; \hat{\overline{\Psi}}_N^{\alpha}
(x_1)\right) | 0 \rangle 
\nonumber\\
&& \times\;\left(i \gamma_{\mu}
\overleftarrow{\partial}^{\mu}_{x_1} + M_N \right)
_{\alpha\;\beta} \; u_N^{\beta} (k_1, \sigma_1) \; {\rm e}^{-i k_1 x_1} \;, 
\label{eq_30}
\end{eqnarray}
where the greek letters $\alpha, \beta$ are Dirac indices. 
The normalization of the nonperturbative QCD vacuum is 
$\langle 0 | 0\rangle =1$,  
and the normalization for the nucleon state reads $\langle N (k_2, \sigma_2) |
N (k_1, \sigma_1) \rangle = 2 E_{k_1} (2 \pi)^3 \delta^{(3)} 
({\bf k}_1-{\bf k}_2) \;\delta_{\sigma_1 \sigma_2}$, 
where $E_{k_1} = \sqrt{{\bf k}_1^2 + M_N^2}$.
Throughout the paper we take the sum convention: If two Dirac 
(or later color) indices are equal or not given explicitly,  
then a sum over them is implied.
The four-momenta are on-shell, $k_1^2 = k_2^2 = M_N^2$;  
for noninteracting nucleons the bare nucleon mass equals 
the physical nucleon mass, $M_N = 938$ MeV. 
The field $\hat{\overline{\Psi}}_N (x_1)$ is the 
interacting (adjoint) nucleon field operator. i.e. off-shell. 
The equal-time anticommutator for the 
interacting nucleon field operator is the same as for the free fields  
and reads 
\begin{eqnarray}
\bigg[\hat{\Psi}^{\alpha}_N ({\bf r}_1, t) , 
\hat{\Psi}^{\beta \; \dagger}_N ({\bf r}_2, t) 
\bigg]_{+} = \delta^{(3)} ({\bf r}_1 - {\bf r}_2) \; 
\delta^{\alpha \beta} \;.
\label{eq_32}
\end{eqnarray}
For the wave function renormalization constant we have 
$0 \le Z_{\Psi}^{-1/2} \le 1$. The free nucleon spinor satisfies  
$(\gamma^{\mu} k_{\mu} - M_N) u_{N} (k, \sigma) = 0$, 
with normalization
$\overline{u}_N (k, \sigma_2) u_N (k, \sigma_1) = 2 M_N\,
\delta_{\sigma_1 \sigma_2}$. 
The operator $\hat{\cal O}$ is, for physical reasons, assumed to  
consist of an even number of fermionic fields, i.e. a bosonic operator, 
according to which the Wick time-ordering,  
${\rm T}_W \hat{A} (x_1) \hat{B} (x_2) = \hat{A} (x_1) \hat{B} (x_2) 
\Theta (t_1 - t_2) + \hat{B} (x_2) \hat{A} (x_1) \Theta (t_2 - t_1)$, 
has been taken in Eq.~(\ref{eq_30}). 

We approximate Eq.~(\ref{eq_30}) by introducing a noninteracting nucleon
field operator given by
\begin{eqnarray}
\hat{\Psi}_N^{\alpha} (x) = \int \frac{d^3 {\bf k}}{(2 \pi)^3}
\frac{1}{2 E_k} \sum\limits_{\sigma=1}^2 
\bigg( \hat{a}_{N} (k, \sigma) \; 
u_{N}^{\alpha} (k, \sigma) \;{\rm e}^{- i k x} 
\nonumber\\
+ \hat{b}^{\dagger}_{N} (k, \sigma) \; 
v_{N}^{\alpha} (k, \sigma) \;{\rm e}^{i k x}\bigg)\;, 
\label{appendixB_5}
\end{eqnarray}
with the corresponding anticommutator relations in momentum space  
\begin{eqnarray}
&& \bigg[\hat{a}_{N} (k_1, \sigma_1) ,  
\hat{a}_{N}^{\dagger} (k_2, \sigma_2) \bigg]_{+} = 
\bigg[\hat{b}_{N} (k_1, \sigma_1)  , 
\hat{b}_{N}^{\dagger} (k_2, \sigma_2) \bigg]_{+} 
\nonumber\\
&& = 2 \, E_{k_1} \; (2 \pi)^3 \; \delta^{(3)} ({\bf k}_1 - {\bf k}_2) \; 
\delta_{\sigma_1\,\sigma_2} \;.
\label{appendixB_10}
\end{eqnarray}
Accordingly, $| N(k, \sigma) \rangle = \hat{a}^{\dagger}_N (k, \sigma) 
| 0 \rangle$. For the noninteracting nucleon field operator 
$Z_{\Psi}^{-1/2} = 1$,  
and the equation of motion follows from (\ref{appendixB_5}),  
$\left(i \gamma_{\mu}\overrightarrow{\partial}^{\mu}_x - 
M_N \right) \hat{\Psi}_N (x) = 0$, and for the adjoint noninteracting 
nucleon field operator it reads $\hat{\overline{\Psi}}_N (x)    
\left(i \overleftarrow{\partial}^{\mu}_x \gamma_{\mu} 
+ M_N \right) = 0$, respectively. Then one arrives at 
\begin{eqnarray}
\langle N (k_2, \sigma_2) | \hat{\cal O} (x) | N (k_1, \sigma_1) \rangle
= \int d^4 x_1 \;   
{\rm e}^{- i k_1 x_1}\; \delta(t - t_1) \nonumber\\ 
\times \; \langle N (k_2, \sigma_2) | \bigg[ \hat{\cal O} (x) , 
\hat{\overline{\Psi}}_N^{\alpha} (x_1) \bigg ]_{-} | 0 \rangle 
\;\left(\gamma_0\right)_{\alpha \beta} \; 
u_{N}^{\beta} (k_1, \sigma_1)\;.
\nonumber\\
\label{eq_35}
\end{eqnarray}
Applying this procedure on the left nucleon state yields 
\begin{widetext}
\begin{eqnarray}
\langle N (k_2, \sigma_2) | \hat{\cal O} (x) | N (k_1, \sigma_1) \rangle
= \int d^4 x_1 \int d^4 x_2 \; 
{\rm e}^{- i k_1 x_1 }\; {\rm e}^{i k_2 x_2} \; \delta(t - t_1) 
\; \delta(t - t_2) 
\nonumber\\
\hspace{-5.0cm} \times \; \overline{u}_{N}^{\beta_2} 
(k_2, \sigma_2) \left(\gamma_0\right)_{\beta_2 \; \alpha_2} \; 
\langle 0 | \Bigg[ \hat{\Psi}_N^{\alpha_2} (x_2) , 
\bigg[\hat{\cal O} (x) ,
\hat{\overline{\Psi}}_N^{\alpha_1} (x_1) \bigg]_{-}\Bigg]_{+} 
|0\rangle \; 
\left(\gamma_0\right)_{\alpha_1\;\beta_1} u_{N}^{\beta_1}
(k_1, \sigma_1)\;,
\label{eq_40}
\end{eqnarray}
\end{widetext}
which is symmetric under the replacement 
$\langle 0 | [\hat{\Psi}, [\hat{\cal O},
\hat{\overline \Psi}]_{-}]_{+} | 0 \rangle \rightarrow \langle 0 | [ [
\hat{\Psi} , \hat{{\cal O}} ]_{-} , \hat{\overline \Psi} ]_{+}
| 0 \rangle$. 
Eq.~(\ref{eq_40}) is the central point of our investigation and we call it
nucleon formula. 
This formula resembles the soft pion theorem given in 
Appendix~\ref{sec:chapter7}.
It is worth to underline that, due to the $\delta$-functions in (\ref{eq_40}), 
only the equal-time commutator and anticommutator occur. 
The anticommutator comes into due to the fact that the commutator
in (\ref{eq_35}) between the operator $\hat{\cal O}$ 
(consisting of an even number of fermionic operators) 
and the (adjoint) fermionic field operator $\hat{\overline{\Psi}}_N$
yields an operator consisting of an odd number of 
fermionic field operators. Therefore, when applying LSZ (cf. Eq.~(\ref{eq_30})) 
on the other nucleon state a Dirac time-ordering,
$T_D \hat{A} (x_1) \hat{B} (x_2) = \hat{A} (x_1) \hat{B} (x_2) 
\Theta (t_1 - t_2) - \hat{B} (x_2) \hat{A} (x_1) \Theta (t_2 - t_1)$, 
is needed.

The nucleon formula is valid for a noninteracting nucleon with finite mass 
$M_N$ and finite three momentum ${\bf k}$, and in this respect it goes beyond the 
soft pion 
theorem, which is valid for pions with vanishing four-momentum only. 
In the next section we supplement the nucleon formula with a nucleon field operator 
expressed by quark fields, which allows then the algebraic 
evaluation of bare nucleon matrix elements of quark operators. 

We note a remarkable advantage of the algebraic approach. 
The operator $\hat{\cal O}$ is a composite 
operator, i.e. a product of field operators taken at the same space-time point.
As it stands, such a composite operator needs to be renormalized. 
Therefore, a renormalization $\hat{\cal O}^{\rm ren} = 
\hat{\cal O} - \langle \hat{\cal O} \rangle_0$ 
(we abbreviate $\langle \hat{\cal O} \rangle_0 \equiv 
\langle 0 | \hat{\cal O} | 0 \rangle$), which applies for products of 
noninteracting field operators, has to be implemented \cite{Zuber}. 
However, the term $\langle \hat{\cal O} \rangle_0$
is a c-number and, according to Eq.~(\ref{eq_40}), does not 
contribute because of $[\langle \hat{\cal O} \rangle_0 , 
\hat{\overline{\Psi}}_N]_{-} = 0$. 
Another kind of renormalization for products of interacting field operators is 
also based on subtracting of c-numbers (so called renormalization constants) 
which vanish when applying the commutator in the nucleon formula. 
Therefore, one may consider the composite operator $\hat{\cal O} (x)$ in 
Eq.~(\ref{eq_40}) as a renormalized operator. This feature is also known 
within PCAC and PCBC algebra, and in particular within the soft pion theorem.

\section{\label{sec:chapter3}Choice of nucleon field operator}

The nucleon formula (\ref{eq_40}) can directly be applied 
on a local operator $\hat{\cal O}$ which consists of bare nucleonic degrees of 
freedom, e.g. $\hat{\cal O} = \hat{\overline \Psi}_N \, \hat{\Psi}_N, 
\hat{\overline \Psi}_N \,\gamma_{\mu} \, \hat{\Psi}_N$, etc. 
However, we are interested in operators basing on quark degrees of freedom, 
e.g. $\hat{\cal O} = \hat{\overline q} \, \hat{q},
\hat{\overline q} \,\gamma_{\mu} \, \hat{q}$, etc. 
To make the relation~(\ref{eq_40}) applicable for such cases  
one needs to decompose the bare nucleon field operator $\hat{\Psi}_N$ into 
the three valence quarks, yielding a composite operator $\hat{\psi}_N$ 
to be specified by now \cite{footnote3}. 
Although there has been considerable 
success in understanding the properties of nucleons on the basis of their 
quark substructure as derived within QCD, a rigorous use of QCD for the 
nucleons is not yet in reach. Therefore, in order to gain a nucleon 
field operator which shows up the main features (quantum numbers) of the 
bare nucleon a more phenomenological approach on the basis of the 
quark-diquark picture of baryons \cite{diquarks} is used. 

To be specific we consider the proton. The bare proton state 
$| u u d\rangle$ is defined by the 
$SU(2)$ flavor, $SU(2)$ spin wavefunction of three valence  
quarks. In the quark-diquark model of the bare proton two of these 
valence quarks are regarded as a composite colored particle (diquark)  
which obeys the Bose statistic and which has a mass of the 
corresponding meson (e.g. for QCD with $N_c = 2$ Pauli-G\"ursey symmetry  
\cite{Pauli_Gursey}), i.e. a
mass which is significantly larger than the current quark mass. 
Within quark degrees of freedom the general expression for such a diquark 
can be written as \cite{pcbc7, Pisarski} 
\begin{eqnarray}
\hat{\Phi}^{{\rm a} \, {\rm b}}_{q_1 q_2} (x) 
= \hat{q}_1^{{\rm a} \; {\rm T}} (x) \; C \, \Gamma 
\; \hat{q}_2^{\rm b} (x) \;.
\label{diquark_5}
\end{eqnarray}
Here, $\hat{q}_1^{\rm a}$ and $\hat{q}_2^{\rm b}$ are quark field operators 
of flavor $u$ or $d$ with color index a and b, respectively. 
Throughout the paper all quark 
field operators are solutions of the full Dirac equation, 
$(i \gamma^{\mu} \hat{D}_{\mu} - m_q) \hat{q} (x) = 0$ with
$\hat{D}_{\mu} = \partial_{\mu} - i g_s \, \hat{A}^a_{\mu} {\lambda}^a/2$, 
where $\hat{A}^a_{\mu}$ are the gluon fields and 
${\rm Tr} (\lambda^a \lambda^b) = 2 \delta^{a b}$ 
($a, b = 1, ..., 8$ are Gell-Mann indices, which should 
not be confused with the  
color indices (in roman style) ${\rm a}, {\rm b}, {\rm c}$ 
(later also ${\rm i}\,...,{\rm n}$) of quark fields).  
The equal-time anticommutator for these interacting quark fields is the same as
for free quark fields and reads  
\begin{eqnarray}
[\hat{q}^{\rm a}_{\alpha} ({\bf r}_1 , t) ,
\hat{q}_{\beta}^{{\rm b}\;\dagger} ({\bf r}_2 , t)]_{+} =
\delta^{(3)} \, ({\bf r}_1 - {\bf r}_2) \;\delta_{\alpha \beta} \; 
\delta^{{\rm a}\, {\rm b}}
\;.
\label{quarks}
\end{eqnarray}
The charge conjugation matrix is $C=i \gamma_0 \gamma_2$, and $\Gamma = 
\{ \, {\bf 1}\hspace{-4pt}1, \gamma_5, \gamma_{\mu}, \gamma_{\mu} \gamma_5, 
\sigma_{\mu \nu} \, \}$ is an element of the Clifford algebra. 
$C$ changes the parity of $\Gamma$, e.g. $C \gamma_5$ has positive parity. 
The diquark (\ref{diquark_5}), considered as a composite operator made of 
quark fields, does generally not commute with quark field operators. 
On the other side, if the diquark is
regarded as an effective boson, it commutes with the fermionic quark fields. 
This feature of the diquark, considered as a bosonic quasiparticle, can   
be retained on quark level when 
neglecting the quantum corrections for the quark fields which are participants
of the diquark. Accordingly, the diquark is separated into a classical 
part and a quantum correction 
\begin{eqnarray}
\hat{\Phi}^{{\rm a}\,{\rm b}}_{q_1 \, q_2} (x) = 
q_1^{{\rm a}\, {\rm T}} (x) \; C \; \Gamma\; q_2^{{\rm b}} (x)
+ \Delta \hat{\Phi}^{{\rm a}\,{\rm b}}_{q_1 \, q_2} (x) \;.
\label{eq_diquark_6}
\end{eqnarray}
The classical Dirac spinors $q_1^{\rm a}, q_2^{\rm b}$ are solutions of 
the full Dirac equation $(i {\gamma^{\mu} D_{\mu}} - m_q) q (x) = 0$. 
The classical part in (\ref{eq_diquark_6}), $\Phi^{{\rm a}\,{\rm b}}_{q_1 q_2} = 
q_1^{{\rm a}\, {\rm T}}\,C \,\Gamma\; q_2^{{\rm b}}$,  
commutes with quark field operators. 
To specify the diquark relevant for a proton we note that there are 
only two structures, $\Gamma = \gamma_5$ and $\Gamma = \gamma_5 \gamma_0$, 
which have positive parity and vanishing total spin, 
$J^P = 0^+$ \cite{Pisarski}. This is in line with \cite{interpolating_1}, 
where it was found that the proton has indeed 
a large overlap with the interpolating field $\hat{\eta}_p 
= \epsilon^{{\rm a}\, {\rm b}\, {\rm c}} 
\left(\hat{u}^{{\rm a}\;{\rm T}} C \gamma_5 \, \hat{d}^{\rm b} \right) 
\hat{u}^{\rm c}$, where $\epsilon^{{\rm a}\, {\rm b}\, {\rm c}}$ is the 
total antisymmetric tensor. 
We also remark that in lattice calculations the field $\hat{\eta}_p$ is 
usually used \cite{lattice_2,lattice} since this interpolating field has an 
appropriate nonrelativistic limit.  
In addition, the field $\hat{\eta}_p$ is also a part of the so called 
Ioffe interpolating field, which for the proton is given by 
$\hat{\eta}_{\rm Ioffe} =
2 \epsilon^{{\rm a}\, {\rm b}\, {\rm c}}
\left((\hat{u}^{{\rm a}\;{\rm T}} C \, \hat{d}^{\rm b})
\gamma_5 \hat{u}^{\rm c} - (\hat{u}^{{\rm a}\;{\rm T}} C \gamma_5 \,
\hat{d}^{\rm b}) \hat{u}^{\rm c}\right)$ \cite{ioffe1}. The Ioffe interpolating 
field is usually used in QCD sum rule evaluations; for a more detailed 
motivation see also \cite{ioffe2}.   
These properties in mind we take $\hat{\eta}_p$ 
as a guide for constructing a proton field 
operator and obtain a semiclassical interpolating proton field by neglecting 
the quantum correction of the diquark. Further, we assume that any quark of 
the nucleon can either be a participant of the diquark or can be located 
outside the diquark. 
In this line only two different structures for a semiclassical interpolating 
proton field may occur and, according to this, the general semiclassical field 
operator for a bare proton is a linear combination of both of them: 
\begin{eqnarray}
\hat{\psi}_p^{\alpha} (x) = \epsilon^{{\rm a}\,{\rm b}\,{\rm c}} \;
\bigg[ A_p \left( u^{{\rm a}\;{\rm T}} (x) \; C \, \gamma_5 \; d^{\rm b} (x)
\right) \; \hat{u}^{{\rm c}\; \alpha} (x)
\nonumber\\
+ B_p \left( u^{{\rm a}\;{\rm T}} (x) \; C \, \gamma_5 \; u^{\rm b} (x) \right)
\; \hat{d}^{{\rm c} \; \alpha} (x) \bigg] \;.
\label{diquark_20}
\end{eqnarray}
The colorless operator (\ref{diquark_20}) 
leads to the quantum numbers of a proton (charge, parity, spin, isospin). 
In the following we evaluate proton matrix elements on quark level 
by means of the nucleon formula (\ref{eq_40}) where the field operator 
$\hat{\Psi}_p (x)$ is replaced by $\hat{\psi}_p (x)$ given in 
Eq.~(\ref{diquark_20}). 

Before going further we have to comment 
on the normalization of the field operator 
(\ref{diquark_20}), i.e. on the determination of the  
coefficients $A_p$ and $B_p$ which, in general, are complex quantities. 
With the relativistic normali\-zation for the nucleon state 
and taking into account that there are two u quarks inside the proton we demand 
\cite{numberoperator}
$\langle p(k, \sigma_1) | \hat{u}^{{\rm i}\;\dagger}_{\alpha} (0)\,
\hat{u}^{\rm i}_{\alpha} (0) | p(k, \sigma_2) \rangle = 4 \; E_{k} \;
\delta_{\sigma_1 \sigma_2}$. 
Evaluating this term using the nucleon formula (\ref{eq_40}) with 
(\ref{diquark_20}) we find the 
normalization, at $x = 0$, to be \cite{footnote4} 
\begin{eqnarray}
|A_p|^2 \; \epsilon^{\rm a b c} \;\epsilon^{{\rm a'} {\rm b'}
{\rm c'}} \,\delta^{{\rm c}\,{\rm c'}} \;
{\overline u}^{{\rm a'}\;{\rm T}} \, C \gamma_5 \,
{\overline d}^{{\rm b'}} \; u^{{\rm a}\;{\rm T}} \, C \gamma_5 \, d^{\rm b} 
= 2 \;.
\label{normalization_2}
\end{eqnarray}
From $\langle p (k, \sigma_1) | \hat{d}^{{\rm i}\;\dagger}_{\alpha} (0)
\hat{d}^{\rm i}_{\alpha} (0) | p (k, \sigma_2) \rangle = 2 E_k \;
\delta_{\sigma_1 \sigma_2}$ we deduce 
\begin{eqnarray}
|B_p|^2\; \epsilon^{\rm a b c} \;\epsilon^{{\rm a'} {\rm b'}
{\rm c'}} \,\delta^{{\rm c}\,{\rm c'}} \; 
{\overline u}^{{\rm a'}\;{\rm T}} \, C \gamma_5 \,
{\overline u}^{{\rm b'}} \; u^{{\rm a}\;{\rm T}} \, C \gamma_5 \, 
u^{\rm b} = 1\;.
\label{normalization_3}
\end{eqnarray}
Formula (\ref{eq_40}) in combination with the field operator 
(\ref{diquark_20}) and the normalizations 
(\ref{normalization_2}) and (\ref{normalization_3}) summarizes our 
propositions made for obtaining bare proton matrix elements for 
quark operators.  

For neutron matrix elements in Eq.~(\ref{eq_40}) we have to insert the
semiclassical field operator for the bare neutron, $\hat{\psi}_n (x)$, which is
achieved from (\ref{diquark_20}) by interchanging
$u \leftrightarrow d$, $\hat{u} \leftrightarrow \hat{d}$ and by the replacements
$A_p \rightarrow A_n$, $B_p \rightarrow B_n$. The corresponding normalizations 
for the bare neutron field operator, i.e. the determination of $A_n$ and $B_n$,  
are obtained from Eqs.~(\ref{normalization_2}) and (\ref{normalization_3}) 
by interchanging the up and down quarks, and $A_p \rightarrow A_n$, 
$B_p \rightarrow B_n$.  

\section{\label{sec:chapter4}Testing the nucleon formula}

In the following we will test the outlined formula, Eq.~(\ref{eq_40}) 
with field operator Eq.~(\ref{diquark_20}), and compare with known 
bare nucleon matrix elements. Throughout the paper we evaluate matrix elements 
of a composite operator $\hat{\cal O} (x)$ at $x=0$ and therefore  
omit the argument $x$ in matrix elements. 

\subsection{\label{sec:chapter4A}Electromagnetic and axial vector current}

The electromagnetic current for the noninteracting pointlike neutron on 
hadronic level is zero, due to the vanishing electric charge of the neutron. 
For the noninteracting pointlike proton it is 
given by $\hat{J}_{\mu}^{\rm em} (x) 
= e_p \hat{\overline\Psi}_p (x) \gamma_{\mu} \hat{\Psi}_p (x)$, 
where the electric charge of proton equals the elementary electric charge, 
$e_p = e$. Now there are two possibilities to evaluate such a matrix element 
on hadronic level: either by means of the algebraic approach 
Eq.~(\ref{eq_40}) and the anticommutator relation (\ref{eq_32}), or 
the usual way by means of the field operator  
Eq.~(\ref{appendixB_5}) and the anticommutator relations (\ref{appendixB_10}).  
In both cases it is straightforward to show that 
$\langle p (k_2, \sigma_2) | \hat{J}_{\mu}^{\rm em} | p (k_1, \sigma_1) 
\rangle =  e_p \, \overline{u}_p (k_2, \sigma_2) \gamma_{\mu} 
u_p (k_1, \sigma_1)$ on effective hadronic level. 
As a first test of the nucleon formula we verify this relation on quark level 
where the electromagnetic current is given by 
$\hat{J}_{\mu}^{\rm em} (x) = \frac{2}{3}\,e\, 
\hat{\overline{u}} (x)\,\gamma_{\mu}\,\hat{u} (x) - \frac{1}{3}\, e \, 
\hat{\overline{d}} (x) \,\gamma_{\mu}\,\hat{d} (x)$.  
Indeed, by using Eqs.~(\ref{eq_application_5}) and (\ref{eq_application_10}) 
from Appendix~\ref{sec:chapter8} for the bare proton we get on quark level  
\begin{eqnarray}
\langle p (k_2, \sigma_2) | \hat{J}^{\rm em}_{\mu} | p (k_1, \sigma_1) \rangle
= \, e_p \, \overline{u}_{p} (k_2, \sigma_2) \, 
\gamma_{\mu}\, u_{p} (k_1, \sigma_1)\;.
\nonumber\\
\label{current_5}
\end{eqnarray}
Similar, for the bare neutron, with Eqs.~(\ref{eq_application_15}) and 
(\ref{eq_application_20}) from Appendix~\ref{sec:chapter8}, 
we obtain on quark level  
$\langle n (k_2, \sigma_2) | \hat{J}^{\rm em}_{\mu} | n (k_1, \sigma_1)
\rangle = 0$. 
Both of these findings are in agreement with the results on effective 
hadronic level. 

Now we look at the axial vector current which on hadronic level for a 
noninteracting pointlike nucleon is given by 
$\hat{A}^{a}_{\mu} (x) = g_A^v \, 
\hat{\overline \Psi}_N (x) \gamma_{\mu} \gamma_5 
\frac{\tau^a}{2} \hat{\Psi}_N (x)$ \cite{formfactor}, 
where $g_A^v$ is the axial charge of a bare nucleon. 
For the moment being in this paragraph up to Eq.~(\ref{eq_application_31}) 
$a = 1, 2, 3$ are isospin indices, and 
$\hat{\Psi}_N$, $|N\rangle$ and $u_N$ are isoduplets. The isospin matrices 
$\tau^a$ coincide with Pauli's spin matrices with normalization 
${\rm Tr} ( \tau^a \, \tau^b ) = 2 \delta^{a\,b}$. 

Similarly to the case of electromagnetic current, there are two possibilities 
to evaluate this matrix element on hadronic level:  
by means of the algebraic approach Eq.~(\ref{eq_40}) and the 
anticommutator relation (\ref{eq_32}), or directly by means of the 
field operator Eq.~(\ref{appendixB_5}) and the anticommutator relations 
(\ref{appendixB_10}). 
In both cases one obtains on effective hadronic level the well known result 
for pointlike nucleons,   
$\langle N (k_2, \sigma_2) | \hat{A}^{a}_{\mu} | N(k_1, \sigma_1) \rangle = 
g_A^{v} \, \overline{u}_N (k_2, \sigma_2) \gamma_{\mu} \gamma_5 
\frac{\tau^a}{2} u_N (k_1, \sigma_1)$. 
We will verify this relation on quark level, where the axial vector  
current is defined as $\hat{A}_{\mu}^{a} (x) =  
(\hat{\overline{u}} (x)\; \hat{\overline{d}} (x)) 
\gamma_{\mu} \gamma_5 \frac{\tau^{a}}{2} (\hat{u} (x) \; \hat{d}(x))^{\rm T}$.  
To operate with matrix elements between either bare proton states or 
bare neutron states we use for the nondiagonal cases $(a = 1, 2)$ the assumed 
isospin symmetry relations, cf. \cite{axial_5},  
$\langle p | \hat{\overline{u}} \, \gamma_{\mu} \gamma_5 \, \hat{d} | n \rangle 
= \, \langle p | \hat{\overline{u}} \,\gamma_{\mu} \gamma_5 \, \hat{u} 
- \hat{\overline{d}} \, \gamma_{\mu} \gamma_5\, \hat{d} | p \rangle$ 
and $\langle n | \hat{\overline{d}} \, \gamma_{\mu} \gamma_5 \, \hat{u} | p  
\rangle = \, \langle n | \hat{\overline{d}}
\, \gamma_{\mu} \gamma_5 \, \hat{d} - \hat{\overline{u}} \,
\gamma_{\mu} \gamma_5\, \hat{u} | n \rangle$ \cite{footnote5}.  
Then, taking the solutions of nucleon formula for two-quark operators, 
Eqs.~(\ref{eq_application_5}), (\ref{eq_application_10}) for proton states, 
and Eqs.~(\ref{eq_application_15}), (\ref{eq_application_20}) 
for neutron states (see Appendix~\ref{sec:chapter8}), 
yields on quark level for the bare 
(isoduplet) nucleon  
\begin{eqnarray}
&&\langle N (k_2, \sigma_2) | \hat{A}_{\mu}^{a} | 
N (k_1, \sigma_1) \rangle 
\nonumber\\
&& = \overline{u}_N (k_2, \sigma_2) \, 
\gamma_{\mu} \gamma_5 \, \frac{\tau^a}{2} \, u_N (k_1, \sigma_1) \;.
\label{eq_application_31}
\end{eqnarray}
Comparision of (\ref{eq_application_31}) with the result on effective hadronic 
level yields for the axial charge $g_A^{v} = 1$, in fair agreement  
with the value $g_A^{v} \simeq 0.84$ deduced from MIT Bag model evaluations and 
neutron $\beta$-decay experiment \cite{footnote6}.   

\subsection{\label{sec:chapter4B}Chiral condensate in nucleon}

The chiral condensate inside the nucleon is related to the pion-nucleon 
sigma term \cite{gasser},  
\begin{eqnarray}
\sigma_N &=& \frac{m_q}{2 M_N} \; 
\langle N_{\rm phys} (k, \sigma) | 
\hat{\overline u} \, \hat{u}  + \hat{\overline d}  \, \hat{d}  | 
N_{\rm phys} (k, \sigma) \rangle \;,
\nonumber\\
\label{chiral_1}
\end{eqnarray}
where $2 \, m_q = m_u + m_d$. 
A typical value for the pion-nucleon sigma term is $\sigma_N = 45$ MeV 
\cite{nuclear_matter_1, nuclear_matter_2}.
The sigma term can be decomposed, according to Eq.~(\ref{bare_nucleon_10}),  
into a valence quark contribution (bare nucleon) and a pion cloud contribution 
(sea quarks and gluons): $\sigma_N = \sigma_N^v + \sigma_N^{\pi}$.
To evaluate $\sigma_N^{v}$ we first consider the u quark chiral condensate 
inside the bare proton. With Eqs.~(\ref{eq_application_5}) and 
(\ref{eq_application_10}) one obtains
\begin{eqnarray}
\langle p (k_2, \sigma_2) | \hat{\overline u}  \, \hat{u}  
| p (k_1, \sigma_1) \rangle &=& 2 \; \overline{u}_p (k_2, \sigma_2) \; 
u_p (k_1, \sigma_1)\;,
\nonumber\\
\langle p (k_2, \sigma_2) | \hat{\overline d}  \, \hat{d}  |
p (k_1, \sigma_1) \rangle 
&=& 1 \; \overline{u}_p (k_2, \sigma_2) \; u_p (k_1, \sigma_1) \;.
\nonumber\\
\label{chiral_25}
\end{eqnarray}
These relations show the momentum and 
spin dependence of the chiral condensate inside the bare proton. 
Of course, for a finite-size nucleon there are additional momentum 
dependences for which is accounted for by nucleon formfactors. 
An application of the nucleon formula to the bare neutron reveals the 
isospin symmetry relations
\begin{eqnarray}
\langle n (k_2, \sigma_2) | \hat{\overline u}  \, \hat{u}  |
n (k_1, \sigma_1) \rangle
&=& \langle p (k_2, \sigma_2) | \hat{\overline d}  \, \hat{d}  |
p (k_1, \sigma_1) \rangle \;,
\nonumber\\
\langle n (k_2, \sigma_2) | \hat{\overline d}  \, \hat{d}  |
n (k_1, \sigma_1) \rangle
&=& \langle p (k_2, \sigma_2) | \hat{\overline u}  \, \hat{u}  |
p (k_1, \sigma_1) \rangle \;. 
\nonumber\\
\label{chiral_27}
\end{eqnarray}
Accordingly, it is only necessary to compare the findings for the proton  
with results reported in the literature. 
For the special case $k_1 = k_2$ and $\sigma_1 = \sigma_2$  
Eq.~(\ref{chiral_25}) simplifies to 
\begin{eqnarray}
\frac{1}{2 M_N} \langle p (k, \sigma) | \hat{\overline u}  \, \hat{u}  
| p (k, \sigma) \rangle &=& 2 \; (2.1) \;, 
\nonumber\\
\nonumber\\
\frac{1}{2 M_N} \langle p (k, \sigma) | \hat{\overline d}  \, \hat{d}  
| p (k, \sigma) \rangle &=& 1 \; (1.4)\;.
\label{chiral_28}
\end{eqnarray}
The parenthesized values are the findings of Ref. \cite{additional_chiral} 
for the valence quark contribution which well agree with our results. 
From (\ref{chiral_28}) and isospin symmetry relations (\ref{chiral_27}) 
one may now deduce the valence quark contribution to the 
nucleon sigma term within the algebraic approach,  
\begin{eqnarray}
\sigma_N^v = \frac{m_q}{2 M_N} \langle N (k, \sigma) | 
\hat{\overline u}  \, \hat{u}  + \hat{\overline d}  \, \hat{d} |
N (k, \sigma) \rangle = 3 \, m_q  \;.
\label{chiral_30}
\end{eqnarray}
We compare this result with Ref. \cite{chiral_estimate}, 
where the valence quark 
contribution to the sigma term
has been estimated to be $\sigma_N^v = \sigma_N / (1 + G_S \, f_{\sigma}^2)$.
By using the given values $G_S = 7.91\,{\rm GeV}^{-2}$ and
$f_{\sigma} = 0.393 \, {\rm GeV}$ one obtains $\sigma_N^v = 20$ MeV.
Accordingly, our result (\ref{chiral_30}) is, 
for $m_q \simeq 7$ MeV, in good numerical agreement with \cite{chiral_estimate}. 

Finally, by assuming that the contribution of the pion cloud for the physical 
proton is the same for the chiral u and d quark condensates one can 
get rid of the term $\sigma_N^{\pi}$ by subtracting the chiral d quark from 
the chiral u quark condensate. That means the following approximation should 
be valid 
\begin{eqnarray}
&&\langle p_{\rm phys} (k, \sigma) | \hat{\overline u}  \,
\hat{u}  - \hat{\overline d}  \, \hat{d}  | p_{\rm phys} (k, \sigma) 
\rangle 
\nonumber\\
&&\simeq \langle p(k, \sigma) | \hat{\overline u}  \,
\hat{u}  - \hat{\overline d}  \, \hat{d}  | p(k, \sigma) \rangle = 2 M_N \;,
\label{eq_application_41}
\end{eqnarray}
where we have used (\ref{chiral_25}). 
Indeed, the result (\ref{eq_application_41}) is in fair agreement with 
$\langle p_{\rm phys} (k, \sigma) | \hat{\overline u} \, \hat{u}  - 
\hat{\overline d}  \, \hat{d}  | p_{\rm phys} (k, \sigma) \rangle  
= 2\,M_N\, (M_{\Xi} - M_{\Sigma}) / m_s = 1.3 \; {\rm GeV}$ 
obtained in \cite{additional_chiral}. 

\subsection{\label{sec:chapter4C}Four-quark condensates}

Four-quark condensates seem to be quite important in predicting the   
properties of light vector mesons within the QCD sum rule method 
\cite{zschocke2}. This is related to the fact that in leading-order 
the chiral condensate is numerically suppressed since it appears 
in a renormalization invariant contribution $m_q\, \langle \hat{\overline q} 
\hat{q}\rangle$. Therefore, the gluon condensate and four-quark condensates 
become numerically more important. However, the numerical values of 
four-quark condensates are poorly known, and up to now 
it remains a challenge to estimate their magnitude in a more reliable way.
Accordingly, the evaluation of four-quark condensates inside the nucleon is an 
important issue. Such quantities have been evaluated 
in \cite{nuclear_matter_2} within the groundstate saturation approximation, 
noting the importance of four-quark condensates also for properties of the nucleon 
within the QCD sum rule approach. 
An attempt to go beyond the groundstate saturation approximation 
has been presented in \cite{four_5}, 
where, by using the Nambu-Jona-Lasinio model and including pions and $\sigma$ 
mesons, correction terms have been obtained. 
Further evaluations of four-quark condensates 
beyond the groundstate saturation approximation have been performed in
\cite{matrixlement_nucleon_5} by using a perturbative chiral quark model for 
describing the nucleons. Later, the results of \cite{matrixlement_nucleon_5} 
have been used for evaluating nucleon parameter at finite density within 
QCD sum rules \cite{matrixlement_nucleon_7}. In \cite{lattice_1, lattice_2}  
lattice evaluations for scalar and traceless four-quark operators 
with non-vanishing twist have been reported. In view of these very few results 
obtained so far further insight into such condensates is desirable. 

Before considering this important issue we notice a general 
decomposition of four-quark condensates. Let ${\hat A}$ and ${\hat B}$ 
two arbitrary two-quark operators.  
Then the nucleon expectation value of $\hat{\cal O}={\hat A} {\hat B}$ can  
be decomposed as \cite{footnote8, footnote7}
\begin{eqnarray}
\langle N_{\rm phys} | {\hat A} {\hat B} | N_{\rm phys} \rangle  
= \langle {\hat A} \rangle_0 
\;\langle N_{\rm phys} | {\hat B} | N_{\rm phys} \rangle  
\nonumber\\
+ \; \langle N_{\rm phys} | {\hat A} | N_{\rm phys} \rangle  
\langle {\hat B} \rangle_0 + \; \langle N_{\rm phys} | {\hat A} {\hat B} |  
N_{\rm phys}\rangle^{\rm C}\;, 
\label{factorization_5}
\end{eqnarray}
where the first two terms refer to the so called factorization approximation, 
while 
the last term is a correction term to the factorization approximation 
and describes the scattering of a nucleon with $\hat{B}$ into  
a nucleon and $\hat{A}$, 
i.e. it is a sum over all connected scattering Feynman diagrams 
$N_{\rm phys} + \hat{B} \rightarrow N_{\rm phys} + \hat{A}$. 
The decomposition (\ref{factorization_5}) is matched with the 
decompositions (\ref{four_5}), (\ref{four_10}) and(\ref{mixed_5}) given below, 
as it is seen in \cite{footnote7} where 
we consider an explicit example for the vector channel.  
The first two terms in (\ref{factorization_5}) scale with $N_c^2$ 
($N_c$ denotes the number of colors, for the moment being 
taken as a free parameter of QCD), 
while 
the correction term scales with $N_c$ \cite{factorization1, factorization2}.
Dividing both sides of (\ref{factorization_5}) by $N_c^2$ one recognizes 
that the last term has to be considered as a correction term of the order 
$1/N_c$ \cite{factorization1, factorization2}. 
That means that a factorization of four-quark operators in a cold medium 
is consistent with the large-$N_c$ limit 
\cite{factorization1, factorization2}, a statement which is also valid in  
vacuum \cite{factorization3}. 

\subsubsection{\label{sec:chapter4C1}Flavor-unmixed four-quark condensates}

We start our investigation with the flavor unmixed four-quark condensates and 
consider the general expression of two 
different kinds of flavor unmixed condensates inside the nucleon, 
namely condensates without and with Gell-Mann matrices $\lambda^a$  
($a=1,...,8$ are the Gell-Mann indices, which should not be confused with 
the color indices (in roman style) ${\rm a}, {\rm b}, {\rm c}$ 
(later also ${\rm i}\,...,{\rm n}$) 
of quark fields) 
\cite{four_5, nuclear_matter_2, footnote9}
\begin{widetext}
\begin{eqnarray}
\langle N_{\rm phys} | \hat{\overline q}  \Gamma_1 \hat{q}  \; 
\hat{\overline q}  \Gamma_2 \hat{q}  | N_{\rm phys} \rangle &=& 
\frac{1}{8} \bigg[ {\rm Tr} \left(\Gamma_1\right) {\rm Tr} 
\left(\Gamma_2\right) - \frac{1}{3} {\rm Tr} \left(\Gamma_1 \Gamma_2 \right)
\bigg] \; \langle \hat{\overline{q}} \hat{q} \rangle_0 \; 
\langle N_{\rm phys} | \hat{\overline{q}} \hat{q} | N_{\rm phys} \rangle  
\nonumber\\
&&\hspace{-4.0cm} + \frac{1}{16} \bigg[ {\rm Tr} (\Gamma_1) {\rm Tr} 
(\gamma^{\mu} \Gamma_2) + {\rm Tr} (\Gamma_2) {\rm Tr} (\gamma^{\mu} \Gamma_1)
- \frac{1}{3} {\rm Tr} (\Gamma_1 \gamma^{\mu} \Gamma_2) 
- \frac{1}{3} {\rm Tr} (\Gamma_2 \gamma^{\mu} \Gamma_1) \bigg] 
\nonumber\\
&&\hspace{-3.0cm} \times \; \langle \hat{\overline q} \hat{q} \rangle_0 \; 
\langle N_{\rm phys} |  
\hat{\overline q} \gamma_{\mu} \hat{q} | N_{\rm phys} \rangle \; + \; 
\langle N_{\rm phys} | \hat{\overline q}  \Gamma_1 \hat{q}  \;
\hat{\overline q}  \Gamma_2 \hat{q}  | N_{\rm phys}\rangle^{\rm C} \;,
\label{four_5}
\\
\nonumber\\
\nonumber\\
\langle N_{\rm phys} | \hat{\overline q}  \Gamma_1 \lambda^a \hat{q}  \;
\hat{\overline q}  \Gamma_2 \lambda^a \hat{q}  | N_{\rm phys} \rangle &=&
- \frac{2}{9} \; {\rm Tr} \left(\Gamma_1 \Gamma_2 \right) \; 
\langle \hat{\overline{q}} \hat{q} \rangle_0 \;
\langle N_{\rm phys} | \hat{\overline{q}} \hat{q} | N_{\rm phys} \rangle
\nonumber\\
&& \hspace{-4.0cm} 
- \frac{1}{9} \bigg[{\rm Tr} (\Gamma_1 \gamma^{\mu} \Gamma_2) + 
{\rm Tr} (\Gamma_2 \gamma^{\mu} \Gamma_1) \bigg] 
\langle \hat{\overline q} \hat{q} \rangle_0 
\; \langle N_{\rm phys} | \hat{\overline q} \gamma_{\mu} \hat{q} 
| N_{\rm phys} \rangle 
+ \; \langle N_{\rm phys} | \hat{\overline q}  \Gamma_1 \lambda^a \hat{q}  \;
\hat{\overline q}  \Gamma_2 \lambda^a \hat{q}  | N_{\rm phys}\rangle^{\rm C}\;,
\label{four_10}
\end{eqnarray}
\end{widetext}
where $\hat{\overline{q}} ... \hat{q}$ is either 
$\hat{\overline{u}} ... \hat{u}$ or $\hat{\overline{d}} ... \hat{d}$ 
(the dots stand for $\Gamma$ or $\Gamma \lambda^a$). 
For the chiral condensate we take $\langle \hat{\overline q} \hat{q} \rangle_0 
= - (0.250 \; {\rm GeV})^3$. 
The decompositions of Eqs.~(\ref{four_5}) and (\ref{four_10}) are related to  
(\ref{factorization_5}) by means of a Fierz rearrangement; an explicit 
example for the vector channel is given in \cite{footnote7}. 
The last term on the right side 
of Eqs.~(\ref{four_5}) and (\ref{four_10}) is a correction 
term \cite{four_5} to the groundstate saturation approximation 
\cite{nuclear_matter_2}, describing the scattering process $N_{\rm phys} 
+ \hat{\overline q} ... \hat{q}\rightarrow  
N_{\rm phys} + \hat{\overline q} ... \hat{q}$. 
To get an idea about the magnitude of these correction terms we consider two 
typical examples. The factorization approximation (\ref{four_5}) 
(i.e. without the correction term) yields for the scalar channel 
$\langle N_{\rm phys} | \hat{\overline q}  \hat{q} \,\hat{\overline q} 
\hat{q} | N_{\rm phys}\rangle = -0.173\, {\rm GeV}^4$. 
In \cite{four_5} the correction term to this groundstate saturation 
approximation has been found to be $\langle N_{\rm phys} | \hat{\overline q}  
\hat{q} \, \hat{\overline q}  \hat{q}  | N_{\rm phys}\rangle^{\rm C}  
= 0.011 \, {\rm GeV^4}$. 
As another example we consider the vector channel with Gell-Mann matrices. 
The factorization approximation (\ref{four_10}) 
(i.e. without the correction term) 
yields $\langle N_{\rm phys} | \hat{\overline q} \gamma_{\mu} \lambda^a 
\hat{q} \;\hat{\overline q}  \gamma^{\mu} \lambda^a \hat{q}  
| N_{\rm phys}\rangle = 0.335 \, {\rm GeV}^4$, while the correction term in 
\cite{four_5} is $\langle N_{\rm phys} | \hat{\overline q} 
\gamma_{\mu} \lambda^a 
\hat{q} \;\hat{\overline q}  \gamma^{\mu} \lambda^a \hat{q}  
| N_{\rm phys}\rangle^{\rm C} = - 0.139 \, {\rm GeV}^4$.  
Accordingly, the correction to the groundstate saturation approximation in the 
scalar channel turns out to be less 
than 10 percent, while in the vector channel 
with Gell-Mann matrices it is about 30 percent. 
As we will see, from (\ref{four_5}) and (\ref{four_10}) the valence quark 
contribution can be extracted in a unique way.

Now we evaluate the valence quark 
contribution of four-quark condensates, and start to consider the u quark 
inside the bare proton. Application of the nucleon formula (\ref{eq_40}) with 
the composite proton field operator (\ref{diquark_20}) yields
\begin{eqnarray}
&&\langle p(k_2, \sigma_2) | \hat{\overline{u}}  \Gamma_1 \hat{u} \;
\hat{\overline{u}}  \Gamma_2 \hat{u}  | p(k_1, \sigma_1) \rangle 
\nonumber\\
&& = 
\overline{u}_p^{\beta_2} (k_2, \sigma_2) \; (\gamma_0)_{\beta_2 \alpha_2} \;
(\gamma_0)_{\alpha_1 \beta_1} \; u_p^{\beta_1} (k_1, \sigma_1)\;
\nonumber\\
&& \times \; (\Gamma_1)^{\alpha \beta} (\Gamma_2)^{\gamma \delta} 
\int d^3 {\bf r}_1\;{\rm e}^{i {\bf k}_1 \, {\bf r}_1} \; \int d^3 {\bf r}_2\;
{\rm e}^{- i {\bf k}_2 \, {\bf r}_2}\;
\nonumber\\
&&\times \; \langle 0| \Bigg[ \hat{\psi}_p^{\alpha_2} ({\bf r}_2 , 0) \,,\bigg[
\hat{\overline{u}}^{\rm i}_{\alpha}  \; \hat{u}^{\rm i}_{\beta}  \;
\hat{\overline{u}}^{\rm j}_{\gamma}  \; \hat{u}^{\rm j}_{\delta}   \; , \;
\hat{\overline{\psi}}_p^{\alpha_1} ({\bf r}_1 , 0) \bigg]_{-} \Bigg]_{+}
| 0 \rangle\;.
\nonumber\\
\label{eq_fourquark_3}
\end{eqnarray}
By inserting the 
expression given in Eq.~(\ref{appendixC_5}) in the Appendix~\ref{sec:chapter8} 
into 
(\ref{eq_fourquark_3}) one obtains for the bare proton 
\begin{widetext}
\begin{eqnarray}
\langle p(k_2, \sigma_2) | \hat{\overline{u}}  \Gamma_1 \hat{u} \; 
\hat{\overline{u}}  \Gamma_2 \hat{u}  | p(k_1, \sigma_1) \rangle 
&=& \frac{1}{6} \; \langle \hat{\overline{u}} \hat{u} \rangle_0 \;
\Bigg( 3 \; {\rm Tr} (\Gamma_1) 
\overline{u}_p (k_2, \sigma_2) \Gamma_2 u_p (k_1, \sigma_1) 
+ 3 \; {\rm Tr} (\Gamma_2) \overline{u}_p (k_2, \sigma_2) 
\Gamma_1 u_p (k_1, \sigma_1) 
\nonumber\\
&& - \overline{u}_p (k_2, \sigma_2) \Gamma_1 
\Gamma_2 u_p (k_1, \sigma_1) 
 - \overline{u}_p (k_2, \sigma_2) \Gamma_2 
\Gamma_1 u_p (k_1, \sigma_1) \Bigg) \;.
\label{eq_fourquark_5}
\end{eqnarray}
\end{widetext}
In a similar way one obtains for the four-quark condensates involving 
Gell-Mann matrices 
\begin{eqnarray}
&&\langle p(k_2,\sigma_2) | \hat{\overline{u}}  \Gamma_1 
\lambda^a \hat{u}  \; \hat{\overline{u}}  \Gamma_2 \lambda^a \hat{u}  
| p(k_1,\sigma_1) \rangle 
\nonumber\\
&& = - \frac{8}{9} \langle \hat{\overline{u}} 
\hat{u} \rangle_0 \Bigg(\overline{u}_p (k_2, \sigma_2) \Gamma_1 \Gamma_2 
u_p (k_1, \sigma_1) 
\nonumber\\
&& + \overline{u}_p (k_2, \sigma_2) \Gamma_2 \Gamma_1 u_p (k_1, \sigma_1) 
\Bigg) \;.
\label{eq_fourquark_10}
\end{eqnarray}
For the d flavor we have
\begin{eqnarray}
&&\langle p(k_2,\sigma_2) | \hat{\overline{d}}  
\Gamma_1 \hat{d}  \; \hat{\overline{d}}  \Gamma_2 \hat{d}
| p(k_1,\sigma_1) \rangle 
\nonumber\\
&& = \frac{1}{2} \; \langle p(k_2,\sigma_2) | \hat{\overline{u}}  
\Gamma_1 \hat{u}  \; \hat{\overline{u}}  \Gamma_2 \hat{u}
| p(k_1,\sigma_1) \rangle \;,
\label{eq_fourquark_11}
\\
&&\langle p(k_2,\sigma_2) | \hat{\overline{d}}  \Gamma_1
\lambda^a \hat{d}  \; \hat{\overline{d}}  \Gamma_2 \lambda^a \hat{d}
| p(k_1,\sigma_1) \rangle  
\nonumber\\
&& = \frac{1}{2} \; \langle p(k_2,\sigma_2) | \hat{\overline{u}}  \Gamma_1
\lambda^a \hat{u}  \; \hat{\overline{u}}  \Gamma_2 \lambda^a \hat{u}
| p(k_1,\sigma_1) \rangle \;.
\label{eq_fourquark_12}
\end{eqnarray}
An analog evaluation of these four-quark operators inside the bare neutron 
gives 
\begin{eqnarray}
\langle n (k_2,\sigma_2) | \hat{\overline{u}}  \Gamma_1  
\hat{u}  \; \hat{\overline{u}}  \Gamma_2 \hat{u}  | n (k_1,\sigma_1) 
\rangle 
\nonumber\\
= \langle p (k_2,\sigma_2) | \hat{\overline{d}}  \Gamma_1
\hat{d}  \; \hat{\overline{d}}  \Gamma_2 \hat{d}  | p (k_1,\sigma_1) 
\rangle \;,
\label{isospin_5}
\end{eqnarray}
\begin{eqnarray}
\langle n (k_2,\sigma_2) | \hat{\overline{u}}  \Gamma_1 \lambda^a 
\hat{u}  \; \hat{\overline{u}}  \Gamma_2 \lambda^a \hat{u}  
| n (k_1, \sigma_1) \rangle 
\nonumber\\
= \langle p (k_2, \sigma_2) | \hat{\overline{d}}  \Gamma_1 
\lambda^a \hat{d}  \; \hat{\overline{d}}  \Gamma_2 \lambda^a \hat{d}  
| p (k_1,\sigma_1) \rangle\;,
\label{isospin_10}
\end{eqnarray}
which, like in the case of chiral condensate, reflect the isospin symmetry.  
By interchanging $\hat{u} \leftrightarrow \hat{d}$ 
on both sides in Eq.~(\ref{isospin_5}) 
and (\ref{isospin_10}) one also gets the d flavor four-quark condensates inside 
neutron. The results (\ref{eq_fourquark_5}) - (\ref{isospin_10}) 
for the four-quark condensates inside the bare nucleon distinguish between 
proton and neutron, and they also contain the dependence  
of these condensates on the momentum of the (pointlike) nucleon. 
Of course, as in the case of two-quark matrix elements, 
for a finite-size nucleon 
there are additional momentum dependences implemented in formfactors. 

We compare now these findings of Eqs.~(\ref{eq_fourquark_5}) - 
(\ref{isospin_10}) with Ref.~\cite{nuclear_matter_2}, i.e. we set $k_1 = k_2$ 
and $\sigma_1 = \sigma_2$, 
and average over proton and neutron to get the nucleon condensates, 
$2 \langle N | \hat{\cal O} | N \rangle = \langle p | \hat{\cal O} | p \rangle 
+ \langle n | \hat{\cal O} | n \rangle$.
First we consider the case of scalar four-quark condensate, i.e.  
$\Gamma_1 = \Gamma_2 = {\bf 1}\hspace{-4pt}1$. Then, for the bare nucleon 
one obtains from Eqs.~(\ref{eq_fourquark_5}), (\ref{eq_fourquark_11}) 
and (\ref{isospin_5})
\begin{eqnarray}
\langle N(k,\sigma) | \hat{\overline{q}}  \hat{q}  \;
\hat{\overline{q}}  \hat{q}  | N(k,\sigma) \rangle 
= \frac{11}{2} \; \langle \hat{\overline{q}} \hat{q} \rangle_0 \; M_N \; ,
\label{eq_fourquark_15}
\end{eqnarray}
while, according to the first term in Eq.~(\ref{four_5}) (the second term 
vanishes in this special case), for the physical nucleon the result
\begin{eqnarray}
&&\langle N_{\rm phys} (k, \sigma) | \hat{\overline{q}}  \hat{q}  \; 
\hat{\overline{q}}  \hat{q}  | N_{\rm phys} (k, \sigma) \rangle
= \frac{11}{6} \; 
\langle \hat{\overline{q}} \hat{q} \rangle_0  \; M_N \; \frac{\sigma_N}{m_q} 
\nonumber\\
&& = \frac{11}{6} \; \langle \hat{\overline{q}} \hat{q} \rangle_0  \; M_N \; 
\bigg( \frac{\sigma_N^v}{m_q} + \frac{\sigma_N^{\pi}}{m_q} \bigg) \; 
\label{eq_fourquark_20}
\end{eqnarray}
has been obtained in \cite{nuclear_matter_2}. 
For the last line of Eq.~(\ref{eq_fourquark_20}) 
we have used the decomposition 
$\sigma_N = \sigma_N^v + \sigma_N^{\pi}$, which has 
already been considered in section~\ref{sec:chapter4B}.
Comparing both results, Eqs.~(\ref{eq_fourquark_15}) and 
(\ref{eq_fourquark_20}), one recognizes, by means of relation 
$\sigma_N^v / m_q = 3$ (cf. \cite{chiral_estimate} and Eq.~(\ref{chiral_30})), 
that the result 
(\ref{eq_fourquark_15}) is nothing else but just the valence quark contribution of the scalar four-quark condensate inside the nucleon; 
it is in agreement with the separated valence quark term of 
Eq.~(\ref{eq_fourquark_20}). 

Due to its importance and its instructive property we will also consider the 
case $\Gamma_1 = {\bf 1}\hspace{-4pt}1, \Gamma_2 = \gamma_{\rho}$.
From Eqs.~(\ref{eq_fourquark_5}), (\ref{eq_fourquark_11}) 
and (\ref{isospin_5}) we find
\begin{eqnarray}
\langle N (k, \sigma) | \hat{\overline q} \hat{q} \; \hat{\overline q} 
\gamma_{\rho} \hat{q} | N (k, \sigma) \rangle 
\nonumber\\
= \frac{5}{4} \langle 
\hat{\overline q} \hat{q} \rangle_0 \; \overline{u}_N (k, \sigma) 
\gamma_{\rho} u_N (k, \sigma) \;,
\label{compare_10}
\end{eqnarray}
which is the contribution of the bare nucleon,
i.e. the valence quark contribution.
To compare it with the corresponding result of Ref. \cite{nuclear_matter_2}
we first deduce from Eq.~(\ref{four_5}) that
\begin{eqnarray}
\langle N_{\rm phys} (k, \sigma) | \hat{\overline q} \hat{q} \; 
\hat{\overline q} \gamma_{\rho} \hat{q} | N_{\rm phys} (k, \sigma) \rangle  
\nonumber\\
= \frac{5}{6} \langle \hat{\overline q} \hat{q} \rangle_0 \; 
\langle N_{\rm phys} (k, \sigma) | \hat{\overline q} \gamma_{\rho} \hat{q}
| N_{\rm phys} (k, \sigma) \rangle \;.
\label{compare_15}
\end{eqnarray}
From (\ref{compare_15}) we have to extract the valence quark contribution
by virtue of Eq.~(\ref{bare_nucleon_10}) (with $Z_N \simeq 1$)
\begin{eqnarray}
&& \langle N_{\rm phys} (k, \sigma) | \hat{\overline q} \gamma_{\rho} \hat{q}
| N_{\rm phys} (k, \sigma) \rangle 
\nonumber\\
&& = \langle N (k, \sigma) | \hat{\overline q}
\gamma_{\rho} \hat{q} | N (k, \sigma) \rangle  
+ \sum\limits_n \phi_n^2 \, \langle N n \pi |
\hat{\overline q} \gamma_{\rho} \hat{q} |  N n \pi \rangle \;.
\label{compare_20}
\nonumber\\
\end{eqnarray}
The first term on the right side, which is in fact an average
over proton and neutron,
is the valence quark term we are interested in, while the second
term is the pion cloud contribution. With isospin invariance one
obtains 
\begin{eqnarray}
&&\langle N (k, \sigma) | \hat{\overline q} \gamma_{\rho} \hat{q} 
| N (k, \sigma) \rangle
\nonumber\\
&& = \frac{1}{2} \bigg( \langle p (k, \sigma) | \hat{\overline u} 
\gamma_{\rho} \hat{u}
| p (k, \sigma) \rangle 
+ \langle p (k, \sigma) | \hat{\overline d} \gamma_{\rho}
\hat{d} | p (k, \sigma) \rangle \bigg) 
\nonumber\\
&& = \frac{1}{2} \bigg(2\; \overline{u}_p (k, \sigma) \gamma_{\rho}
u_p (k, \sigma) + 1 \; \overline{u}_p (k, \sigma) \gamma_{\rho} u_p (k, \sigma) \bigg)
\nonumber\\
&& = \frac{3}{2} \overline{u}_N (k, \sigma) \gamma_{\rho} u_N (k, \sigma) \;,
\label{compare_25}
\end{eqnarray}
where for the last term we have set $u_p = u_N$ because of $M_p = M_N$.
By inserting (\ref{compare_20}) into (\ref{compare_15}) and using  
(\ref{compare_25}) we obtain
\begin{eqnarray}
&&\langle N_{\rm phys} (k, \sigma) | \hat{\overline q} \hat{q} \;
\hat{\overline q} \gamma_{\rho} \hat{q} | N_{\rm phys} (k, \sigma) \rangle  
\nonumber\\
&& = \frac{5}{4} \langle \hat{\overline q} \hat{q} \rangle_0 \;
\overline{u}_N (k, \sigma) \gamma_{\rho} u_N (k, \sigma)
\nonumber\\
&& + \frac{5}{6} \langle \hat{\overline q} \hat{q} \rangle_0  
\sum\limits_n \phi_n^2 \, \langle N n \pi |
\hat{\overline q} \gamma_{\rho} \hat{q}
|  N n \pi \rangle\;.
\label{compare_30}
\end{eqnarray}
The first term on the right side of Eq.~(\ref{compare_30})
agrees with our result (\ref{compare_10}), while the second term 
on the right side of Eq.~(\ref{compare_30})  
is a factorization approximation of the 
expression $\sum\limits_n \phi_n^2 \, \langle N n \pi | 
\hat{\overline q} \hat{q} 
\hat{\overline q} \gamma_{\rho} \hat{q} |  N n \pi \rangle$.  
From that it becomes obvious 
that our result (\ref{compare_10}) is nothing else but
just the valence quark contribution of (\ref{compare_30}). 

Other combinations of Clifford matrices, 
like $\Gamma_1=\gamma_5$ and $\Gamma_2= \gamma_{\rho}$, 
with or without Gell-Mann matrices, can be evaluated and compared in the 
same way \cite{footnote10}.  
This means that within the algebraic approach (\ref{eq_40}) 
for evaluating bare nucleon matrix elements we find an agreement 
for all of the flavor-unmixed four-quark condensates if one takes from the 
corresponding results of Ref.~\cite{nuclear_matter_2} the valence quark 
contribution, for instance by means of the decomposition  
$\sigma_N = \sigma_N^v + \sigma_N^{\pi}$.  
Therefore, one actually may consider our 
evaluation as a re-evaluation of the valence quark 
contribution of the four-quark condensates of Ref.~\cite{nuclear_matter_2}
and a confirmation of their results within an independent microscopic 
approach (quark-diquark picture of the nucleon). 
But we have to be aware that such an agreement between our algebraic 
approach and the factorization approximation applies only for the 
valence quark contribution of nucleon matrix elements. 
Especially, such an agreement is not expected when taking into account 
the pion cloud contributions according to Eq.~(\ref{bare_nucleon_10}).   

Having found the remarkable agreement with valence quark contribution of the 
factorization approximation it becomes interesting to compare our results 
also with other evaluations in the literature. However, it turns out 
that a comparison with the recent lattice data of Ref. \cite{lattice_1}
is quite involved since in \cite{lattice_1} the
condensates have been evaluated at a 
renormalization scale of about $\mu_{\rm lattice}^2 \simeq 5 \; \rm {GeV}^2$, 
which is considerably higher than our renormalization point 
of about $\mu^2 \simeq 1 \; \rm {GeV}^2$  
(our renormalization point is hidden in the chiral condensate,
i.e. $\langle \hat{\overline q} \hat{q}\rangle_0 (\mu^2)$).   
To scale the lattice data from $5 \, {\rm GeV}^2$ down to the hadronic scale 
of $1 \, {\rm GeV}^2$ one needs to know the matrix of anomalous dimension for 
all of the four-quark operators which accounts for the effect of operator 
mixing among the four-quark condensates. This operator mixing may change 
considerably the numerical values and even the signs of the 
four-quark condensates given in \cite{lattice_1}. 
Another method which seems also capable to compare our results
with lattice data would be to scale our renormalization point $\mu^2$
up to the lattice scale $\mu_{\rm lattice}^2$. Such a procedure requires, 
however, the knowledge of renormalization scale dependence of the chiral  
condensate. Alltogether, performing these procedures 
is out of the scope of the present paper and we therefore abandon
a comparison of our results with the ones given in Ref. \cite{lattice_1}.

In view of the mentioned points and especially in view of 
an acceptable lucidity of our paper,  
we restrict ourselves to a comparison with the 
recently obtained four-quark condensates of Ref. \cite{matrixlement_nucleon_5}. 
Due to the specific notation for the four-quark condensates choosen in Ref. 
\cite{matrixlement_nucleon_5} we list our results for the valence quark 
contribution of scalar four-quark condensates for all channels in the same 
way as in Ref. \cite{matrixlement_nucleon_5}. 
Our results for the valence quark contribution can be obtained from 
Eqs.~(\ref{eq_fourquark_5}) - (\ref{isospin_10}) by averaging over proton 
and neutron ($k_1 = k_2, \sigma_1 = \sigma_2$):
\begin{widetext}
\begin{eqnarray}
\langle N | \frac{2}{3} \hat{\overline q} \hat{q} \,
\hat{\overline q} \hat{q}
- \frac{1}{2} \hat{\overline q} \lambda^a \hat{q} \,
\hat{\overline q} \lambda^a \hat{q} | N \rangle
&& = - 0.0733 \; {\rm GeV}^4 \; (- 0.117\;{\rm GeV}^4) \;,\;
\label{our_5}
\\
\langle N | \frac{2}{3} \hat{\overline q} \gamma_5 \hat{q} \,
\hat{\overline q} \gamma_5 \hat{q}
- \frac{1}{2} \hat{\overline q} \gamma_5 \lambda^a \hat{q} \,
\hat{\overline q} \gamma_5 \lambda^a \hat{q} | N \rangle
&& = - 0.0147 \; {\rm GeV}^4 \;(- 0.0567 \;{\rm GeV}^4) \;,\;
\label{our_6}
\\
\langle N | \frac{2}{3}\hat{\overline q} \gamma_{\mu}
\hat{q} \, \hat{\overline q} \gamma^{\mu} \hat{q}
- \frac{1}{2} \hat{\overline q} \gamma_{\mu} \lambda^a \hat{q} \,
\hat{\overline q} \gamma^{\mu} \lambda^a \hat{q} | N \rangle
&& =  - 0.0586 \; {\rm GeV}^4 \;(- 0.0582 \;{\rm GeV}^4) \;, \;
\label{our_7}
\\
\langle N | \frac{2}{3} \hat{\overline q} \gamma_{\mu}
\gamma_5 \hat{q} \, \hat{\overline q} \gamma^{\mu} \gamma_5 \hat{q}
- \frac{1}{2} \hat{\overline q} \gamma_{\mu} \gamma_5 \lambda^a \hat{q} \,
\hat{\overline q} \gamma^{\mu} \gamma_5 \lambda^a \hat{q} | N \rangle
&& = + 0.0586 \; {\rm GeV}^4 \;(+ 0.0567 \;{\rm GeV}^4) \;,\;
\label{our_8}
\\
\langle N | \frac{2}{3} \hat{\overline q} \sigma_{\mu \nu}
\hat{q} \, \hat{\overline q} \sigma^{\mu \nu} \hat{q}
- \frac{1}{2} \hat{\overline q} \sigma_{\mu \nu} \lambda^a
\hat{q} \, \hat{\overline q} \sigma^{\mu \nu}
\lambda^a \hat{q} | N \rangle
&& = - 0.176 \; {\rm GeV}^4 \;(- 0.356 \;{\rm GeV}^4) \;.\; 
\label{our_9}
\end{eqnarray}
\end{widetext}
The values parenthesized are the results for these condensates as given in  
Ref. \cite{matrixlement_nucleon_5}, but for the physical nucleon, 
i.e. for a nucleon which contains the valence quark, sea quark and gluon 
contributions \cite{footnote11}.  
Since we have compared our evaluated valence quark 
contribution with the total contribution for the physical nucleon of
Ref. \cite{matrixlement_nucleon_5} it becomes obvious that one
may actually not expect a perfect numerical agreement.
The more interesting fact is to notice that the valence quark contribution 
for the vector and axial vector channel, (\ref{our_7}) and (\ref{our_8}), 
respectively, agrees very well with the total contribution for the physical 
nucleon.
For the scalar, axial scalar and tensor channel the signs for the 
valence quark contribution and total contribution of Ref. 
\cite{matrixlement_nucleon_5} agree, while the numerical values  
differ. This illustrates 
that the sea quark and gluon contributions are expected 
to give noticeable contributions.  

\subsubsection{\label{sec:chapter4C2}Flavor-mixed four-quark condensates}

For the flavor-mixed four-quark condensates the general decomposition reads
\cite{nuclear_matter_2}
\begin{eqnarray}
&&\langle N_{\rm phys} | \hat{\overline u}  \Gamma_1 \hat{u}  \;
\hat{\overline d}  \Gamma_2 \hat{d}  | N_{\rm phys} \rangle 
\nonumber\\
&& = \frac{1}{8} \langle \hat{\overline q} \hat{q} \rangle_0 \;
\langle N_{\rm phys} | \hat{\overline q} \hat{q} | N_{\rm phys} \rangle 
\;{\rm Tr} \left( \Gamma_1 \right) \, {\rm Tr} \left( \Gamma_2 \right)
\nonumber\\
&&+ \langle \hat{\overline q} \hat{q} \rangle_0 \; \langle N_{\rm phys} |
\hat{\overline q} \gamma_{\mu} \hat{q} | N_{\rm phys} \rangle 
\bigg[ {\rm Tr} \left( \Gamma_1 \right) \, {\rm Tr} \left( \gamma^{\mu}
\Gamma_2 \right)
\nonumber\\
&&+ {\rm Tr} \left( \Gamma_2 \right) \, {\rm Tr} \left( \gamma^{\mu}
\Gamma_1 \right) \bigg] + \langle N_{\rm phys} | \hat{\overline u} 
\Gamma_1 \hat{u}  \;
\hat{\overline d}  \Gamma_2 \hat{d}  | N_{\rm phys}\rangle^{\rm C} \;,
\nonumber\\
\label{mixed_5}
\end{eqnarray}
where the isospin symmetry relations (\ref{chiral_27}) have been used. The 
last term in (\ref{mixed_5}) is a correction term to the factorization 
approximation, describing the scattering process 
$N_{\rm phys} + \hat{\overline d} \Gamma_2 \hat{d} 
\rightarrow N_{\rm phys} + \hat{\overline u} \Gamma_1 \hat{u}$.  
The decomposition (\ref{mixed_5}) is, like (\ref{four_5}) and 
(\ref{four_10}), 
matched with Eq.~(\ref{factorization_5}) by means of a Fierz transformation. 
The flavor-mixed condensates with Gell-Mann matrices vanish 
in the factorization approximation, 
$\langle N_{\rm phys} | \hat{\overline u}  \Gamma_1 \lambda^a \hat{u}  \;
\hat{\overline d}  \Gamma_2 \lambda^a \hat{d}  | N_{\rm phys} \rangle = 0$ 
\cite{nuclear_matter_2} (in \cite{zschocke1} we have found small corrections 
to the factorization approximation of flavor-mixed condensates for the vector 
and axial vector channel with Gell-Mann matrices).

Now we consider the valence quark contribution of the flavor mixed four-quark 
condensates, i.e. the contribution of the bare nucleon. 
Application of our nucleon formula (\ref{eq_40}) with (\ref{diquark_20}) 
yields for the bare proton  
\begin{eqnarray}
&&\langle p(k_2, \sigma_2) | \hat{\overline{u}}  \Gamma_1 \hat{u}  \;
\hat{\overline{d}}  \Gamma_2 \hat{d}  | p(k_1, \sigma_1) \rangle 
\nonumber\\
&& = \frac{1}{4} \Bigg( 2 \;\langle \hat{\overline{u}} \hat{u} \rangle_0 \;  
{\rm Tr} (\Gamma_1) \; \overline{u}_p (k_2, \sigma_2) \, \Gamma_2 \, 
u_p (k_1, \sigma_1) 
\nonumber\\
&& + \, 1 \; \langle \hat{\overline{d}} \hat{d} \rangle_0 \; 
{\rm Tr} (\Gamma_2) \; \overline{u}_p (k_2, \sigma_2) \, \Gamma_1 \, 
u_p (k_1, \sigma_1) \Bigg)\;,
\label{eq_fourquark_25}
\\
&&\langle p(k_2, \sigma_2) | \hat{\overline{u}}  \Gamma_1 \lambda^a 
\hat{u}  \;
\hat{\overline{d}}  \Gamma_2 \lambda^a \hat{d}  
| p(k_1, \sigma_1) \rangle = 0\;,
\label{eq_fourquark_30}
\end{eqnarray}
while for the bare neutron we find 
\begin{eqnarray}
&&\langle n(k_2, \sigma_2) | \hat{\overline{u}}  \Gamma_1 \hat{u}  \;
\hat{\overline{d}}  \Gamma_2 \hat{d}  | n(k_1, \sigma_1) \rangle
\nonumber\\
&& = \frac{1}{4} \Bigg( 1 \;\langle \hat{\overline{d}} \hat{d} \rangle_0 \;
{\rm Tr} (\Gamma_1) \; \overline{u}_n (k_2, \sigma_2) \, \Gamma_2 \, 
u_n (k_1, \sigma_1)
\nonumber\\
&&+ \, 2 \; \langle \hat{\overline{u}} \hat{u} \rangle_0 \;
{\rm Tr} (\Gamma_2) \; \overline{u}_n (k_2, \sigma_2) \, \Gamma_1 \, 
u_n (k_1, \sigma_1) \Bigg)
\;,
\label{eq_fourquark_31}
\\
&&\langle n(k_2, \sigma_2) | \hat{\overline{u}}  \Gamma_1 \lambda^a
\hat{u}  \;
\hat{\overline{d}}  \Gamma_2 \lambda^a \hat{d}
| n(k_1, \sigma_1) \rangle = 0\;.
\label{eq_fourquark_32} 
\end{eqnarray}
The Eqs.~(\ref{eq_fourquark_25}) and (\ref{eq_fourquark_31}) are  
generalized expressions of the factorization approximation given in  
\cite{nuclear_matter_2} because of the dependence on nucleon momentum 
and the distinction between proton and neutron.
The results of Eqs.~(\ref{eq_fourquark_30}) and 
(\ref{eq_fourquark_32}) are in agreement with the factorization approximation 
\cite{nuclear_matter_2}, but, as mentioned above, beyond the factorization 
approximation these condensates are nonvanishing \cite{zschocke1}. 

To compare the obtained results of Eqs.~(\ref{eq_fourquark_25}) and 
(\ref{eq_fourquark_31}) with the factorization approximation, i.e. 
neglecting the correction term in (\ref{mixed_5}),  
we consider the special case $k_1 = k_2$, 
$\sigma_1 = \sigma_2$ and 
$\Gamma_1 = \Gamma_2 = {\bf 1}\hspace{-4pt}1$. 
For the bare nucleon one obtains by averaging over 
the bare proton and bare neutron, 
\begin{eqnarray}
\langle N (k, \sigma) | \hat{\overline{u}}  \hat{u}  \;
\hat{\overline{d}}  \hat{d}  | N (k, \sigma) \rangle =  
6 \; \langle \hat{\overline{q}} \hat{q} \rangle_0 \; M_N \;.
\label{eq_fourquark_35}
\end{eqnarray}
This result has to be compared with the corresponding finding of 
\cite{nuclear_matter_2} which according to Eq. (\ref{mixed_5}) reads
\begin{eqnarray}
&&\langle N_{\rm phys} (k, \sigma) | \hat{\overline{u}}  \hat{u}  \;
\hat{\overline{d}}  \hat{d}  | N_{\rm phys} (k, \sigma) \rangle 
= 2 \; \langle \hat{\overline{q}} \hat{q} \rangle_0 
\; M_N \; \frac{\sigma_N}{m_q} 
\nonumber\\
&& =  2 \; \langle \hat{\overline{q}} \hat{q} \rangle_0 \; M_N \; 
\bigg(\frac{\sigma_N^v}{m_q} + \frac{\sigma_N^{\pi}}{m_q} \bigg) \;,
\label{eq_fourquark_36}
\end{eqnarray}
where in the last expression we have used the decomposition 
$\sigma_N = \sigma_N^v + \sigma_N^{\pi}$. 
By means of the relation $\sigma_N^v / m_q =3$ 
(cf. \cite{chiral_estimate} and Eq.~(\ref{chiral_30})) the result 
(\ref{eq_fourquark_35}) is in agreement with the separated valence quark 
contribution of (\ref{eq_fourquark_36}).  
As in the cases considered in the previous subsection 
\ref{sec:chapter4C1} 
such an agreement with Ref. \cite{nuclear_matter_2} 
can be achieved for all the other combinations of $\Gamma_1$ and $\Gamma_2$ of 
the Clifford algebra. 

As in the case of flavor-unmixed four-quark operators 
we compare our findings for the valence quark contribution of 
flavor-mixed condensates, (\ref{eq_fourquark_25}) and (\ref{eq_fourquark_31}), 
with the total result for the physical nucleon of 
Ref. \cite{matrixlement_nucleon_5} to examine the 
magnitude and sign of our results.  
According to Eqs.~(\ref{eq_fourquark_25}) - (\ref{eq_fourquark_32}) only 
the valence quark contribution of the scalar channel does not vanish. 
Its numerical magnitude 
\begin{eqnarray}
\langle N | \frac{2}{3} \hat{\overline u} \hat{u} \,
\hat{\overline d} \hat{d} 
- \frac{1}{2} \hat{\overline u} \lambda^a \hat{u} \,
\hat{\overline d} \lambda^a \hat{d} | N \rangle
\nonumber\\
= - 0.0586 \; {\rm GeV}^4 \;(- 0.094\; {\rm GeV}^4) \;, 
\label{mixed_15}
\end{eqnarray}
turns out to be comparable with the evaluation of 
\cite{matrixlement_nucleon_5} for the total contribution of the 
physical nucleon given in the parentheses in (\ref{mixed_15}). The numerical 
difference in magnitude is caused by sea quark and gluon contributions. 

To summarize this section, we have evaluated the valence quark contribution 
to flavor-unmixed and flavor-mixed four-quark condensates for the 
u  and d flavor inside proton and neutron. The results for the 
flavor-unmixed operators are given by the Eqs.~(\ref{eq_fourquark_5})
- (\ref{isospin_10}), and
the results for the flavor-mixed operators are given by the 
Eqs.~(\ref{eq_fourquark_25}) - (\ref{eq_fourquark_32}). 
We have seen that our findings for the four-quark
condensates within the algebraic approach, which is by far a different
method than the used one of Ref.~\cite{nuclear_matter_2}, are in agreement
with the large-$N_c$ limit \cite{factorization1,factorization2} and with the
results of Ref.~\cite{nuclear_matter_2} when taking from there the
valence quark contribution only.
It seems admissible, especially in view of the agreement with valence quark 
contribution of factorization, that the nucleon formula yields reliable
results for the valence quark contribution of four-quark condensates
inside the nucleon. 

\section{\label{sec:chapter5}Six-quark Condensates}

Six-quark condensates become important mainly for two reasons. First, 
in the operator product expansion (OPE) of current correlators one usually 
takes into account
all terms up to the order of the four-quark condensates and
neglects the contributions of higher order. Such an approximation may work or
may not work, depending on the specific physical system under consideration.
Accordingly, one has to be aware about the contribution of the next order to
decide how good such an approximation is. This is also necessary for
the more involved case of finite density, where a Gibbs average over all 
hadronic states of the correlator under consideration has to be taken. 
Indeed, a very recent estimate of such higher contributions beyond the 
four-quark condensate for the nucleon correlator in matter underlines also the 
importance of an estimate for the six-quark condensates inside the nucleon 
\cite{hungary}. And second, it is 
well known that instantons give rise to corrections to the Wilson coefficients 
of six-quark condensates \cite{six_quark_1, six_quark_2}. These corrections 
cause a substantial enhancement of the vacuum contribution of six-quark 
condensates in the OPE of current correlators. 
To investigate such current correlators at finite density implies 
the knowledge of the nucleon matrix elements of six-quark condensates.  
Here, after getting confidence on our proposed approach in the previous 
sections, we will use the nucleon formula to evaluate the valence quark 
contribution of six-quark condensates inside the nucleon. 

Within our algebraic approach by using the nucleon formula (\ref{eq_40}) 
with (\ref{diquark_20}) we obtain for the u flavor inside the bare proton 
\begin{eqnarray}
&&\hspace{0.0cm}
\langle p (k_2, \sigma_2) | \hat{\overline u} \, \Gamma_1 \, \hat{u} \;
\hat{\overline u}  \, \Gamma_2 \, \hat{u} \;
\hat{\overline u} \, \Gamma_3 \, \hat{u} | p (k_1, \sigma_1) \rangle 
\nonumber\\
&&\hspace{-0.2cm}
= \overline{u}_p^{\beta_2} (k_2, \sigma_2) \; (\gamma_0)_{\beta_2 \alpha_2} \;
(\gamma_0)_{\alpha_1 \beta_1} \; u_p^{\beta_1} (k_1, \sigma_1)
\nonumber\\
&&\hspace{-0.2cm}\times \; \left(\Gamma_1\right)^{\alpha \beta} \; 
\left(\Gamma_2\right)^{\gamma \delta} \;
\left(\Gamma_3\right)^{\epsilon \zeta} 
\int d^3 {\bf r}_1 \; 
{\rm e}^{i {\bf k}_1 \, {\bf r}_1}
\; \int d^3 {\bf r}_2\;
{\rm e}^{- i {\bf k}_2 \, {\bf r}_2}\;
\nonumber\\
&&\hspace{-0.2cm}
\langle 0| \Bigg[ \hat{\psi}_p^{\alpha_2} ({\bf r}_2 , 0) \,,\bigg[
\hat{\overline{u}}^{\rm i}_{\alpha}  \, \hat{u}^{\rm i}_{\beta}  \,
\hat{\overline{u}}^{\rm k}_{\gamma}  \, \hat{u}^{\rm k}_{\delta}  \,
\hat{\overline{u}}^{\rm m}_{\epsilon}  \, \hat{u}^{\rm m}_{\zeta}  \, , \,
\hat{\overline{\psi}}_p^{\alpha_1} ({\bf r}_1 , 0) \bigg]_{-} \Bigg]_{+}
| 0 \rangle\,.
\nonumber\\
\label{six_5}
\end{eqnarray}
The commutator-anticommutator is given in Eq.~(\ref{six_10})  
in Appendix~\ref{sec:chapter8} for a more general 
case. According to this result the six-quark condensate inside the 
bare proton is reduced to a four-quark condensate in vacuum. We note one of 
these four-quark condensates in vacuum saturation approximation \cite{sumrule} 
\begin{eqnarray}
&&\langle \hat{u}^{\rm j}_{\beta} \; \hat{\overline u}^{\rm k}_{\gamma}\;
\hat{u}^{\rm l}_{\delta} \; \hat{\overline u}^{\rm m}_{\epsilon}  \rangle_0 
= \frac{1}{(12)^2} 
\langle \hat{\overline u} \hat{u} \rangle_0^2 \; 
\nonumber\\
&&\times\; \bigg( 
\delta_{\beta \gamma} \delta_{\delta \epsilon} \; \delta^{{\rm j}\, {\rm k}} 
\delta^{{\rm l}\, {\rm m}} 
- \delta_{\beta \epsilon} \delta_{\gamma \delta} \;
\delta^{{\rm j}\, {\rm m}} \delta^{{\rm k}\, {\rm l}}\bigg) \;.
\label{six_15}
\end{eqnarray}
When evaluating all of the four-quark condensates of Eq.~(\ref{six_10}) 
in the same way one obtains, by using the normalization 
(\ref{normalization_2}), 
\begin{widetext}
\begin{eqnarray}
\langle p (k_2, \sigma_2) | \hat{\overline u} \, \Gamma_1 \, \hat{u} \;
\hat{\overline u}  \, \Gamma_2 \, \hat{u} \;
\hat{\overline u} \, \Gamma_3 \, \hat{u} | p (k_1, \sigma_1) \rangle
&=& \frac{2}{(12)^2} \; \langle \hat{\overline u} \hat{u} \rangle_0^2  
\nonumber\\
&&\hspace{-7.50cm}\times\; \bigg[
\overline{u}_p (k_2, \sigma_2) \left(\Gamma_1 \, \Gamma_2 \, \Gamma_3 
+ \Gamma_1 \, \Gamma_3 \, \Gamma_2
+ \Gamma_2 \, \Gamma_1 \, \Gamma_3 \right) u_p (k_1, \sigma_1)
+ \overline{u}_p (k_2, \sigma_2) \left(\Gamma_2 \, \Gamma_3 \, \Gamma_1
+ \Gamma_3 \, \Gamma_1 \, \Gamma_2
+ \Gamma_3 \, \Gamma_2 \, \Gamma_1 \right) u_p (k_1, \sigma_1)
\nonumber\\
&&\hspace{-7.50cm}- 3 \, \overline{u}_p (k_2, \sigma_2) 
\Gamma_1 \, \Gamma_3
u_p (k_1, \sigma_1) \; 
\;{\rm Tr} \left( \Gamma_2 \right) 
- 3 \, \overline{u}_p (k_2, \sigma_2)
\Gamma_2 \, \Gamma_3 
u_p (k_1, \sigma_1) \; {\rm Tr} \left( \Gamma_1 \right) 
- 3 \, \overline{u}_p (k_2, \sigma_2) \Gamma_1 \, \Gamma_2  
u_p (k_1, \sigma_1)\;{\rm Tr} \left( \Gamma_3 \right)
\nonumber\\
&&\hspace{-7.50cm}- 3 \, \overline{u}_p (k_2, \sigma_2)
\Gamma_3 \, \Gamma_2 u_p (k_1, \sigma_1) \;{\rm Tr} \left( \Gamma_1 \right)
 - 3 \, \overline{u}_p (k_2, \sigma_2)
\left(\Gamma_2 \, \Gamma_1 \right) 
u_p (k_1, \sigma_1)
\;{\rm Tr} \left( \Gamma_3 \right)
\nonumber\\
&& \hspace{-7.50cm} - 3 \, \overline{u}_p (k_2, \sigma_2)
\left(\Gamma_3 \, \Gamma_1 \right) 
u_p (k_1, \sigma_1) \;{\rm Tr} \left( \Gamma_2 \right)
+ 9 \, \overline{u}_p (k_2, \sigma_2) \Gamma_3  
u_p (k_1, \sigma_1)
\;{\rm Tr} \left( \Gamma_1 \right) \;{\rm Tr} \left( \Gamma_2 \right) 
\nonumber\\
&& \hspace{-7.50cm} 
+ 9 \, \overline{u}_p (k_2, \sigma_2) \Gamma_2 u_p (k_1, \sigma_1)
\;{\rm Tr}\left(\Gamma_1 \right) \;{\rm Tr} \left( \Gamma_3 \right)
+ 9 \, \overline{u}_p (k_2, \sigma_2) \Gamma_1 
u_p (k_1, \sigma_1) 
\;{\rm Tr}\left( \Gamma_2 \right) \;{\rm Tr}\left( \Gamma_3 \right)
\nonumber\\
&& \hspace{-7.50cm} - 3 \, \overline{u}_p (k_2, \sigma_2) \Gamma_3  
u_p (k_1, \sigma_1) \;{\rm Tr} \left( \Gamma_1 \Gamma_2 \right)
- 3 \, \overline{u}_p (k_2, \sigma_2) 
\Gamma_2 u_p (k_1, \sigma_1) \;{\rm Tr} \left(\Gamma_1 \Gamma_3 \right) 
- 3 \, \overline{u}_p (k_2, \sigma_2) \Gamma_1 
u_p (k_1, \sigma_1) \;{\rm Tr} 
\left( \Gamma_2 \Gamma_3 \right) \bigg]\;. 
\label{six_16}
\end{eqnarray}
\end{widetext}
For the d flavor six-quark condensate we get 
\begin{eqnarray}
\langle p (k_2, \sigma_2) | \hat{\overline d} \, \Gamma_1 \, \hat{d} \;
\hat{\overline d}  \, \Gamma_2 \, \hat{d} \;
\hat{\overline d} \, \Gamma_3 \, \hat{d} | p (k_1, \sigma_1) \rangle 
\nonumber\\
= \frac{1}{2}
\langle p (k_2, \sigma_2) | \hat{\overline u} \, \Gamma_1 \, \hat{u} \;
\hat{\overline u}  \, \Gamma_2 \, \hat{u} \;
\hat{\overline u} \, \Gamma_3 \, \hat{u} | p (k_1, \sigma_1) \rangle\;.
\label{six_17}
\end{eqnarray}
These findings are, to the best of our knowledge, the first attempts 
to estimate the size of six-quark condensates inside a (bare) nucleon. 
An analog evaluation for the bare neutron reveals the 
isospin symmetry relations  
\begin{eqnarray}
\langle n (k_2, \sigma_2) | \hat{\overline u} \, \Gamma_1 \, \hat{u} \;
\hat{\overline u}  \, \Gamma_2 \, \hat{u} \;
\hat{\overline u} \, \Gamma_3 \, \hat{u} | n (k_1, \sigma_1) \rangle 
\nonumber\\
= \langle p (k_2, \sigma_2) | \hat{\overline d} \, 
\Gamma_1 \, \hat{d} \;
\hat{\overline d}  \, \Gamma_2 \, \hat{d} \;
\hat{\overline d} \, \Gamma_3 \, \hat{d} | p (k_1, \sigma_1) \rangle \;,
\label{six_18}
\\
\langle n (k_2, \sigma_2) | \hat{\overline d} \, \Gamma_1 \, \hat{d} \;
\hat{\overline d}  \, \Gamma_2 \, \hat{d} \;
\hat{\overline d} \, \Gamma_3 \, \hat{d} | n (k_1, \sigma_1) \rangle
\nonumber\\
= \langle p (k_2, \sigma_2) | \hat{\overline u} \, 
\Gamma_1 \, \hat{u} \;
\hat{\overline u}  \, \Gamma_2 \, \hat{u} \;
\hat{\overline u} \, \Gamma_3 \, \hat{u} | p (k_1, \sigma_1) \rangle \;.
\label{six_19}
\end{eqnarray}
Finally, by averaging over the proton matrix elements, 
Eqs.~(\ref{six_16}) and (\ref{six_17}), and the neutron matrix elements, 
Eqs.~(\ref{six_18}) and (\ref{six_19}), one 
gets the six-quark condensates inside a bare nucleon.  
For instance, the six-quark condensate for the scalar channel is  
found to be 
\begin{eqnarray}
\langle N(k, \sigma) | \hat{\overline q}\, \hat{q} \; 
\hat{\overline q}\, \hat{q} 
\; \hat{\overline q} \,\hat{q} | N(k, \sigma) \rangle 
= \frac{55}{8} \langle \hat{\overline q} \hat{q} \rangle_0^2 \; M_N\;.
\label{example_six} 
\end{eqnarray} 

Further, we present results for six-quark condensates which contain 
Gell-Mann matrices. Note that only nucleon matrix 
elements of colorless operators do not vanish, i.e. only two 
Gell-Mann matrices may occur. With the general result (\ref{six_10})
in Appendix~\ref{sec:chapter8} 
and normalization (\ref{normalization_2}) we obtain
\begin{widetext}
\begin{eqnarray}
&&\langle p (k_2, \sigma_2) | \hat{\overline u} \, \Gamma_1 \lambda^{a} \, 
\hat{u} \; \hat{\overline u}  \, \Gamma_2 \lambda^{a} \, \hat{u} \;
\hat{\overline u} \, \Gamma_3 \, \hat{u} | p (k_1, \sigma_1) \rangle
= \frac{2}{(12)^2} \; \langle \hat{\overline u} \hat{u} \rangle_0^2 \; 
\nonumber\\
&&\hspace{-0.8cm}\times\;
\Bigg[\frac{16}{3}  
\overline{u}_p (k_2, \sigma_2) \left(\Gamma_1 \, \Gamma_2 \, \Gamma_3
+ \Gamma_1 \, \Gamma_3 \, \Gamma_2
+ \Gamma_2 \, \Gamma_1 \, \Gamma_3 \right) u_p (k_1, \sigma_1)
 + \frac{16}{3} 
\overline{u}_p (k_2, \sigma_2) \left(\Gamma_2 \, \Gamma_3 \, \Gamma_1
+ \Gamma_3 \, \Gamma_1 \, \Gamma_2
+ \Gamma_3 \, \Gamma_2 \, \Gamma_1 \right)  u_p (k_1, \sigma_1)
\nonumber\\
&&\hspace{-0.5cm} - 16 \, \overline{u}_p (k_2, \sigma_2) \Gamma_2 \Gamma_1\,
\overline{u}_p (k_1, \sigma_1) \; {\rm Tr} \left( \Gamma_3\right)
-16 \, \overline{u}_p (k_2, \sigma_2) \Gamma_1 \Gamma_2\,
\overline{u}_p (k_1, \sigma_1) \; {\rm Tr} \left( \Gamma_3 \right) 
 - 16 \, \overline{u}_p (k_2, \sigma_2) ¸\Gamma_3\,
\overline{u}_p (k_1, \sigma_1) \; {\rm Tr} \left( \Gamma_1 \Gamma_2 \right)
\Bigg] \;. 
\nonumber\\
\label{six_20}
\end{eqnarray}
\end{widetext}
For the d flavor we get
\begin{eqnarray}
\langle p (k_2, \sigma_2) | \hat{\overline d} \, \Gamma_1 \lambda^{a} \,
\hat{d} \; \hat{\overline d}  \, \Gamma_2 \lambda^{a} \, \hat{d} \;
\hat{\overline d} \, \Gamma_3 \, \hat{d} | p (k_1, \sigma_1) \rangle 
\nonumber\\
= \frac{1}{2}\;
\langle p (k_2, \sigma_2) | \hat{\overline u} \, \Gamma_1 \lambda^{a} \,
\hat{u} \; \hat{\overline u}  \, \Gamma_2 \lambda^{a} \, \hat{u} \;
\hat{\overline u} \, \Gamma_3 \, \hat{u} | p (k_1, \sigma_1) \rangle\;. 
\label{six_25}
\end{eqnarray}
Finally, evaluating these operators inside the bare neutron we obtain 
the isospin symmetry relations 
\begin{eqnarray}
\langle n (k_2, \sigma_2) | \hat{\overline u} \, \Gamma_1 \lambda^{a} \,
\hat{u} \; \hat{\overline u}  \, \Gamma_2 \lambda^{a} \, \hat{u} \;
\hat{\overline u} \, \Gamma_3 \, \hat{u} | n (k_1, \sigma_1) \rangle
\nonumber\\
= \langle p (k_2, \sigma_2) | \hat{\overline d} \, \Gamma_1 \lambda^{a} \,
\hat{d} \; \hat{\overline d}  \, \Gamma_2 \lambda^{a} \, \hat{d} \;
\hat{\overline d} \, \Gamma_3 \, \hat{d} | p (k_1, \sigma_1) \rangle\;,
\label{six_30}
\\
\langle n (k_2, \sigma_2) | \hat{\overline d} \, \Gamma_1 \lambda^{a} \,
\hat{d} \; \hat{\overline d}  \, \Gamma_2 \lambda^{a} \, \hat{d} \;
\hat{\overline d} \, \Gamma_3 \, \hat{d} | n (k_1, \sigma_1) \rangle
\nonumber\\
= \langle p (k_2, \sigma_2) | \hat{\overline u} \, \Gamma_1 \lambda^{a} \,
\hat{u} \; \hat{\overline u}  \, \Gamma_2 \lambda^{a} \, \hat{u} \;
\hat{\overline u} \, \Gamma_3 \, \hat{u} | p (k_1, \sigma_1) \rangle\;.
\label{six_35}
\end{eqnarray}
As before, by averaging over the proton matrix elements, 
Eqs.~(\ref{six_20}) and (\ref{six_25}), 
and the neutron matrix elements, 
Eqs.~(\ref{six_30}) and (\ref{six_35}), one obtains 
the six-quark condensates containing Gell-Mann matrices inside a bare nucleon. 
The presented findings for six-quark condensates inside the 
bare nucleon provide basic results for further investigations   
beyond the order of four-quark condensates within the QCD sum rule approach 
for the nucleon in matter (for the nucleon sum rule in vacuum see  
\cite{ioffe1, ioffe2}, and for the nucleon sum rule in matter see  
\cite{nuclear_matter_1, nuclear_matter_2, nuclear_matter_3}).  
Investigations aiming at predictions beyond the order of 
four-quark condensates, however, 
imply in addition to the evaluation of the six-quark condensates also the 
knowledge of the Wilson coefficients for all of these six-quark condensates. 
So far, these coefficients in the OPE for the nucleon correlator have been 
determined up to the order of the four-quark condensates 
\cite{hungary}.

\section{\label{sec:chapter6}Summary}

An algebraic approach for evaluating bare nucleon matrix elements has been 
presented. The supposed nucleon formula (\ref{eq_40}) 
relates bare nucleon matrix elements 
to vacuum matrix elements and, 
therefore, new parameters for evaluating them are not needed. 
A feature of the algebraic method is that the valence quark contribution of  
two-quark, four-quark and six-quark condensates 
can be evaluated on the same footing.
One aim of the present paper is to demonstrate how the nucleon formula works 
and to test it in several cases. In doing so, 
the nucleon has been considered as a composite pointlike object, 
described by a valence quark and a valence diquark 
approximated by two classical Dirac spinors. Neither sea quarks nor gluons, 
or in a hadronic language no meson cloud, have been taken into account here. 
Accordingly, the results presented have to be considered as pure valence 
quark contribution to the matrix elements under consideration. 

A consideration of the electromagnetic current (\ref{current_5}) 
and the axial vector 
current (\ref{eq_application_31}) 
for the bare nucleon reveals the expected current 
structure for a pointlike object. We have evaluated then the valence quark 
contribution of the chiral condensate (\ref{chiral_30}), 
finding the relation $\sigma_N^v = 3 \,m_q$  
which turns out to be in numerical agreement with results obtained in  
an earlier work \cite{chiral_estimate}. 

Furthermore, the 
valence quark contribution of four-quark condensates has been investigated. 
Our results are given in Eqs.~(\ref{eq_fourquark_5}) - (\ref{isospin_10}) 
for flavor-unmixed operators, and in Eqs.~(\ref{eq_fourquark_25}) - 
(\ref{eq_fourquark_32}) for flavor-mixed operators.  
In the special case $k_1 = k_2,\sigma_1 = \sigma_2$ we find an interesting 
agreement with the groundstate saturation approximation explored  
in Ref. \cite{nuclear_matter_2} if one separates  
the valence quark contribution of four-quark condensates from that results. 
In this respect our approach yields 
an independent re-evaluation and confirmation of the results of 
Ref. \cite{nuclear_matter_2}, because 
both methods are different from the conceptional point of view, which is 
already interesting in itself. Even more, our algebraic approach presented 
recovers the dependence of condensates on momentum for a pointlike nucleon and 
distinguishes between proton and neutron matrix elements. 
In this respect it goes beyond the common factorization approximation. 
In this context 
it is worth to underline that the agreement between our algebraic 
approach and the groundstate saturation approximation has been found for the 
bare nucleon, and not for the physical nucleon. 
In Eqs.~(\ref{our_5}) - (\ref{our_9}) and (\ref{mixed_15})
we have compared our results with values 
of four-quark condensates inside the physical nucleons recently obtained 
within a chiral quark model \cite{matrixlement_nucleon_5}. 

As a further application of nucleon formula we have presented results for 
six-quark condensates inside the bare nucleon, given in 
Eqs.~(\ref{six_16}) - (\ref{six_19}) and 
Eqs.~(\ref{six_20}) - (\ref{six_35}), respectively. 
These values obtained are, 
to the best of our knowledge, the first 
evaluation of six-quark condensates inside (bare) nucleons. 
Finally, in Eqs.~(\ref{chiral_30}), (\ref{eq_fourquark_15}) and 
(\ref{example_six}) we have given more explicit examples for the 
scalar channel of  
two-quark, four-quark and six-quark operators, respectively, 
inside the bare nucleon, showing up an interesting alternative change in 
the algebraic sign from two-quark to four-quark and from four-quark 
to six-quark condensates.  

A remark should also be in order about nucleon matrix elements of gluonic
operators. Evaluating such operators within the algebraic approach presented
requires, in general, the implementation of gluonic degrees of freedom 
into the composite nucleon field operator (\ref{diquark_20}). 
Such an implementation  
might be provided by the quark-gluon interpolating nucleon field
discussed in another context in \cite{schaefer}. 

The algebraic approach can be extended into several directions. 
First, the description of the proton   
core with the field operator (\ref{diquark_20}), and the corresponding one for 
the bare neutron, can be improved, e.g.  
by implementing an effective potential between the valence quarks. 
And second, the pion cloud of nucleon, accounting for virtual  
sea quarks and gluons inside the physical nucleon, can be
implemented within the Tamm-Dancoff method. To get an algebraic approach also
for such a case one has to combine the nucleon formula (\ref{eq_40}) with the
soft pion theorem (\ref{appendixB_25}). This implies the evaluation of the 
coefficients $\phi_n$ in (\ref{bare_nucleon_5}) within the renormalizable 
pion-nucleon interaction Hamiltonian, which is therefore a topic of further 
investigations. Accordingly, for the time being the 
application of nucleon formula (\ref{eq_40}) in combination with the 
field operator (\ref{diquark_20}) has to be considered as a first step 
in determining more accurately nucleon matrix elements of quark operators. 
In summary, we arrive at the conclusion that a reliable evaluation of 
quark operators inside nucleons can be based on a purely algebraic approach. 
This triggers the hope that predictions of in-medium properties of hadrons 
become more precise in future. 

\section{Acknowledgement} 

The work is supported by BMBF and GSI. 
We would like to thank Prof. Leonid P. Kaptari, Dr. Hanns-Werner Barz,
Dr. Vahtang Gogohia, Dr. Ralf Sch\"utzhold, Prof. Vladimir Shabaev, 
Dr. Gyuri Wolf and Dr. Miklos Z\'et\'enyi for useful discussions.  
The authors also thank Prof. Horst St\"ocker and Prof. Laszlo P. Csernai
for their kind support during the work.
One of the authors (S.Z.) thanks for the warm hospitality at 
the Frankfurt Institute for Advanced Studies (FIAS) 
in Frankfurt a.M./Germany and at the Research Institute for  
Particle and Nuclear Physics (KFKI-RMKI) in Budapest/Hungary, 
and  he would also like to express his 
gratitude to Iris Zschocke and Steffen K\"ohler for their
enduring encouragement during the work.

\appendix

\section{\label{sec:chapter7}Soft pion theorem}

In this Appendix we recall a soft pion theorem relevant 
for our purposes to show the similarity in derivation and final form of it 
with the nucleon formula 
(\ref{eq_40}). Let us consider the general pion matrix element of an 
operator $\hat{\cal O} (x)$ which in general may depend on space and time
\cite{hosaka}
\begin{eqnarray}
&&\langle \pi^b (p_2) | \hat{\cal O} (x) | \pi^a (p_1) \rangle 
\nonumber\\ 
&& = i Z_{\varphi}^{-1/2}
\int d^4 x_1 {\rm e}^{- i p_1 x_1} \left( \Box_{x_1} + m_{\pi}^2 \right)
\nonumber\\
&&\times \;\langle \pi^b (p_2) |\; {\rm T}_W
\left( \hat{\cal O} (x) \; \hat{\varphi}^a (x_1)
\right)
| 0 \rangle\;,
\label{appendixA_5}
\end{eqnarray}
where the LSZ reduction formalism has been applied on pion state
$| \pi^a (p_1) \rangle$.
Here, the letters $a, b = 1, 2, 3$ denote isospin indices. 
The normalization of pion state is $\langle \pi^b (p_2) | \pi^a (p_1) \rangle
= 2 E_{p_1} (2 \pi)^3 \delta^{(3)} ({\bf p}_1-{\bf p}_2)\; \delta^{a b}$, 
where $ E_{p_1}=\sqrt{{\bf p}_1^2 + m_{\pi}^2}$.
The wave function renormalization
constant is $0 \le Z_{\varphi}^{-1/2} \le 1$. The
normalization of nonperturbative QCD vacuum is $\langle 0 | 0\rangle =1$.
Here, $x_1=({\bf r}_1, t_1)$ is the space-time 
four-vector, and $T_W$ denotes the Wick time
ordering.
The states $| \pi^a (p_i) \rangle$ are, of course, on-shell states, i.e. 
solutions of the Klein-Gordon equation for noninteracting pions, while the
field operator $\hat{\varphi}^a$, in general,
is the interacting field, i.e. it is off-shell.
The four momenta in (\ref{appendixA_5}) are on-shell,  
i.e.  $p_1^2 = p_2^2 = m_{\pi}^2$.
The soft pion theorem is valid for a noninteracting pion field
(i.e. $Z_{\varphi}^{-1/2} =1$), wich is a solution of the Klein-Gordon equation 
$\left( \Box_{x} + m_{\pi}^2 \right) \hat{\varphi}^a (x)$.
In order to be complete in the representation we will also specify
the noninteracting pion field operator
\begin{eqnarray}
\hat{\varphi}^a (x) = \int \frac{d^3 {\bf p}}{(2 \pi)^3}
\frac{1}{2 E_p}
\bigg( \hat{a}^a (p) \; {\rm e}^{- i p x} \; + \;
\hat{b}^{a \; \dagger} (p) \; {\rm e}^{i p x} \bigg) \;,
\nonumber\\
\label{pion_5}
\end{eqnarray}
where the creation and annihilation operators obey 
the following commutator relations
\begin{eqnarray}
\bigg[\hat{a}^{a} (p_1) \; , \; \hat{a}^{b \; \dagger} (p_2) \bigg]_{-}
= \bigg[\hat{b}^{a} (p_1) \; , \; \hat{b}^{b \; \dagger} (p_2) \bigg]_{-}
\nonumber\\
= 2 E_{p_1} \; (2 \pi)^3 \; \delta^{(3)} ({\bf p}_1 - {\bf p}_2) \;
\delta^{a b}.
\label{pion_10}
\end{eqnarray}
Accordingly, $| \pi^{a} (p) \rangle = \hat{a}^{a\;\dagger} (p) | 0 \rangle$.
From (\ref{pion_5}) and (\ref{pion_10}) we deduce the equal-time commutator
for the noninteracting pion fields,
\begin{eqnarray}
\bigg[ \hat{\varphi}^{a} ({\bf r}_1 , t) \; , \;
\partial_0 \, \hat{\varphi}^{b} ({\bf r}_2 , t) \; \bigg]_- =
\, i \; \delta^{(3)} ({\bf r}_1 - {\bf r}_2) \; \delta^{a b} \;.
\label{pion_1}
\end{eqnarray}
In addition, the soft pion theorem is only valid in case of
vanishing four-momentum $p_1^{\mu} \rightarrow 0$ which implies
$m_{\pi} = 0$. Then we get
\begin{eqnarray}
&&\lim_{\atop p_1 \to 0}
\langle \pi^b (p_2) | \hat{\cal O} (x) | \pi^a (p_1) \rangle  
\nonumber\\
&& = i \int d^4 x_1\; \Box_{x_1} \;
\langle \pi^b (p_2) | \;{\rm T}_W \left( \hat{\cal O} (x) \; 
\hat{\varphi}^a (x_1)
\right)
| 0 \rangle
\nonumber\\
&& = - i \langle \pi^b (p_2) | \bigg[\hat{\cal O} (x) \; , \;
\frac{\partial}{\partial t_1} \hat{\varphi}^{a} (x_1) \bigg]_{-}
|0\rangle \; \delta(t - t_1)\;.
\label{appendixA_10}
\nonumber\\
\end{eqnarray}
In the last line we have used the equation of motion for the massless pion
field, i.e. $\Box_{x_1} \;\hat{\varphi}^a (x_1) = 0$.
Now, the PCAC hypothesis, which asserts a relation between the axial current 
and the pion field ($f_{\pi} \simeq 92.4$ MeV is the pion decay constant), 
\begin{eqnarray}
\hat{A}_{\mu}^a (x) = - f_{\pi} \partial_{\mu} \hat{\varphi}^a (x)\;,
\label{appnedixA_15}
\end{eqnarray}
is inserted into Eq.~(\ref{appendixA_10})  
(by means of the field equation for the noninteracting pion
it becomes obvious from (\ref{appnedixA_15}) that in the limit of
vanishing pion mass the PCAC goes over to a conserved axial vector current
(see also \cite{formfactor})). The  
axial vector current $\hat{A}_{\mu}^a (x)$ obeys 
the well known current algebra commutation relations \cite{lit4} 
which directly leads to the soft pion theorem relevant for our purposes 
\cite{hosaka}
\begin{eqnarray}
\lim_{\atop p_1 \to 0}
\langle \pi^b (p_2) | \hat{\cal O} (x) | \pi^a (p_1) \rangle =
\frac{i}{f_{\pi}} \; \int d^4 x_1\; 
\nonumber\\
\times \; \langle \pi^b (p_2) |
\bigg[\hat{\cal O} (x) \; , \; \hat{A}_0^a (x_1) \; \bigg]_{-} | 0 \rangle
\; \delta(t - t_1)\;.
\label{appendixB_20}
\end{eqnarray}
One may apply the same steps as before on the other pion state
as well, ending up with the soft pion theorem 
\begin{eqnarray}
&&\lim_{\atop p_2 \to 0} \lim_{\atop p_1 \to 0}
\langle \pi^b (p_2) | \hat{\cal O} (x) | \pi^a (p_1) \rangle 
\nonumber\\ 
&& = \frac{1}{f_{\pi}^2} \; \int d^4 x_1\; \int d^4 x_2\; \delta(t - t_1) \;
\delta(t - t_2) 
\nonumber\\
&&\times \; \langle 0 | \Bigg[ \hat{A}_0^b (x_2) \; , \; \bigg[
\hat{\cal O} (x) \; , \; \hat{A}_0^a (x_1) \bigg]_{-} \Bigg]_{-} | 0 \rangle
\nonumber\\
&& = \frac{1}{f_{\pi}^2} \langle 0 | \Bigg[ \hat{Q}_A^b ,
\bigg[ \hat{\cal O} (x) \; , \; \hat{Q}_A^a
\bigg]_{-} \Bigg]_{-} |0\rangle\;.
\label{appendixB_25}
\end{eqnarray}
where $\hat{Q}_A^a$ is the (time independent) axial charge, 
$\hat{Q}_A^a = \int d^3 {\bf r} \; \hat{A}_0^a ({\bf r}, t)$.
It seems expedient to emphasize that the QCD quark degrees of
freedom were not necessary for deriving the soft pion theorem.
If one expresses the axial vector current by quark fields,
$\hat{A}_{\mu}^a (x) =
\hat{\overline{\Psi}} (\tau^a/2) \gamma_{\mu} \gamma_5 \hat{\Psi}$
with $\hat{\Psi} = (\hat{u}\; \hat{d})^{\rm T}$, and the pion fields
as well by means of their interpolating fields 
(i.e. composite quark fields which have the quantum numbers of pions), 
then the relation
(\ref{appnedixA_15}), the current algebra and, therefore, the soft pion 
theorem (\ref{appendixB_25}) 
can also be established within QCD degrees of freedom.
This theorem can then also be used for evaluating pion matrix elements 
of quark operators (see, for instance, \cite{Hatsuda, pion}). 
Summarizing, the soft pion theorem is valid for a noninteracting
pion field with vanishing pion four-momentum. In many applications
such a restriction is not problematic since the pion mass is small compared 
to a typical hadronic scale of about 1 GeV. 

\section{\label{sec:chapter8}Evaluating Eq.~(\ref{six_5})}

To evaluate Eq.~(\ref{six_5}) we start with the case of two quark field 
operators. The nucleon formula (\ref{eq_40}) with (\ref{diquark_20}) yields 
\begin{eqnarray}
&&\langle p (k_2, \sigma_2) | \hat{\overline{u}}_{\alpha}^{\rm i}  \; 
\hat{u}^{\rm i}_{\beta}  | p (k_1, \sigma_1) \rangle 
\nonumber\\
&& = \overline{u}_{p}^{\beta_2} (k_2, \sigma_2) \, 
(\gamma_0)_{\beta_2 \, \alpha_2} \;
(\gamma_0)_{\alpha_1 \, \beta_1} \; u_{p}^{\beta_1} (k_1, \sigma_1) 
\nonumber\\
&&\times \; \int d^3 {\bf r}_1 \; \int d^3 {\bf r}_2 \; 
{\rm e}^{i {\bf k}_1 {\bf r}_1} \; {\rm e}^{- i {\bf k}_2 {\bf r}_2} \;
\nonumber\\
&&\times \; \langle 0 | \Bigg[ \hat{\psi}^{\alpha_2}_p ({\bf r}_2, 0) , \bigg[ 
\hat{\overline{u}}_{\alpha}^{\rm i}  \; \hat{u}^{\rm i}_{\beta}  , 
\hat{\overline{\psi}}^{\alpha_1}_p ({\bf r}_1 , 0) \bigg]_{-} \Bigg]_{+} | 0 
\rangle \;.
\label{B_5}
\end{eqnarray}
Inserting the proton field operator (\ref{diquark_20}) and using 
$[\hat{A} \hat{B} , \hat{C}]_{-} = \hat{A} [\hat{B} , \hat{C}]_{+} - 
[\hat{A} , \hat{C}]_{+} \hat{B}$ for the commutator we obtain  
\begin{eqnarray}
\bigg[
\hat{\overline{u}}_{\alpha}^{\rm i}  \; \hat{u}^{\rm i}_{\beta}  ,
\hat{\overline{\psi}}^{\alpha_1}_p ({\bf r}_1 , 0) \bigg]_{-} 
&=& A_p^{*} \; 
\epsilon^{{\rm a}\, {\rm b}\, {\rm c}} \; 
\bigg(\overline{u}^{{\rm a}\;{\rm T}} ({\bf r}_1) 
\; C \gamma_5 \; \overline{d}^{\rm b} ({\bf r}_1) \bigg) 
\nonumber\\ 
&&\times \; \delta^{{\rm i}\, {\rm c}} \; 
(\gamma_0)_{\beta \alpha_1} \; \delta^{(3)} ({\bf r}_1) \; 
\hat{\overline u}^{\rm i}_{\alpha}  \;.
\label{B_10}
\end{eqnarray}
In the same way we get  
\begin{eqnarray}
&&\Bigg[ \hat{\psi}^{\alpha_2}_p ({\bf r}_2, 0) , \bigg[
\hat{\overline{u}}_{\alpha}^{\rm i}  \; \hat{u}^{\rm i}_{\beta}  ,
\hat{\overline{\psi}}^{\alpha_1}_p ({\bf r}_1 , 0) \bigg]_{-} \Bigg]_{+}
\nonumber\\
&&= |A_p|^2 \; \epsilon^{{\rm a}\, {\rm b}\, {\rm c}} \; 
\epsilon^{{\rm a'}\, {\rm b'}\, {\rm c'}}  
(\gamma_0)_{\beta \alpha_1} \; (\gamma_0)_{\alpha \alpha_2}  
\nonumber\\
&& \times \; \delta^{{\rm c}\, {\rm c'}} 
\; \delta^{(3)} ({\bf r}_1) \; \delta^{(3)} ({\bf r}_2) 
\nonumber\\
&&\times \; 
\left(u^{{\rm a'}\;{\rm T}}
({\bf r}_2) \, C \gamma_5 \, d^{\rm b'} ({\bf r}_2) \right)\;
\left(\overline{u}^{{\rm a}\;{\rm T}} 
({\bf r}_1) \, C \gamma_5 \,\overline{d}^{\rm b} ({\bf r}_1) 
\right) \;.
\nonumber\\
\label{B_15}
\end{eqnarray}
Integrating over both delta-functions and then using the 
normalization (\ref{normalization_2}) we obtain  
\begin{eqnarray}
&&\int d^3 {\bf r}_1 \; {\rm e}^{i {\bf k}_1 {\bf r}_1}  \int d^3 {\bf r}_2 \; 
{\rm e}^{- i {\bf k}_2 {\bf r}_2} \; 
\nonumber\\
&&\times \; \langle 0 | \Bigg[ \hat{\psi}^{\alpha_2}_p ({\bf r}_2, 0) , \bigg[
\hat{\overline{u}}_{\alpha}^{\rm i}  \; \hat{u}^{\rm i}_{\beta}  ,
\hat{\overline{\psi}}^{\alpha_1}_p ({\bf r}_1 , 0) \bigg]_{-} \Bigg]_{+} 
| 0 \rangle 
\nonumber\\
&& = 2 \; (\gamma_0)_{\beta \alpha_1} \; 
(\gamma_0)_{\alpha \alpha_2} \;,
\label{B_20}
\end{eqnarray}
and with (\ref{B_5}) 
\begin{eqnarray}
\langle p (k_2, \sigma_2) | \hat{\overline{u}}_{\alpha}^{\rm i}  \; 
\hat{u}^{\rm i}_{\beta}  | p (k_1, \sigma_1) \rangle 
= 2 \; \overline{u}_p^{\alpha} (k_2, \sigma_2) \; u_p^{\beta} 
(k_1, \sigma_1)\;.
\label{eq_application_5}
\nonumber\\
\end{eqnarray}
We note an analog relation for the d quark  
\begin{eqnarray}
\langle p (k_2, \sigma_2) | \hat{\overline{d}}_{\alpha}^{\rm i}  \; 
\hat{d}^{\rm i}_{\beta} 
| p (k_1, \sigma_1) \rangle &=& 1 \; 
\overline{u}_p^{\alpha} (k_2, \sigma_2) \; u_p^{\beta} (k_1, \sigma_1)
\;,
\nonumber\\
\label{eq_application_10}
\end{eqnarray}
while for the neutron we have
\begin{eqnarray}
\langle n (k_2, \sigma_2) | \hat{\overline{u}}_{\alpha}^{\rm i}  \; 
\hat{u}^{\rm i}_{\beta}  | n (k_1, \sigma_1) \rangle &=& 1 \; 
\overline{u}_n^{\alpha} (k_2, \sigma_2) \; u_n^{\beta} (k_1, \sigma_1)
\;,
\nonumber\\
\label{eq_application_15}
\\
\langle n (k_2, \sigma_2) | \hat{\overline{d}}_{\alpha}^{\rm i}  \; 
\hat{d}^{\rm i}_{\beta}  | n (k_1, \sigma_1) \rangle &=& 2\; 
\overline{u}_n^{\alpha} (k_2, \sigma_2) \; u_n^{\beta} (k_1, \sigma_1)
\;.
\nonumber\\
\label{eq_application_20}
\end{eqnarray}
Using relations like $[\hat{A}, \hat{B} \hat{C}]_+ = 
[\hat{A}, \hat{B}]_+ \hat{C} - \hat{B} [\hat{A}, \hat{C}]_- $ 
analog equations for the four-quark condensates can be obtained. 
Two illustrative examples are given 
for the flavor-unmixed four-quark condensate 
for the proton  
\begin{eqnarray}
&&\int d^3 {\bf r}_1\;{\rm e}^{i {\bf k}_1 
\, {\bf r}_1} \; \int d^3 {\bf r}_2\; 
{\rm e}^{- i {\bf k}_2 \, {\bf r}_2}\; 
\nonumber\\
&&\times \; \langle 0| \Bigg[ \hat{\psi}_p^{\alpha_2} ({\bf r}_2 , 0) \,,\bigg[ 
\hat{\overline{u}}^{\rm i}_{\alpha}  \; \hat{u}^{\rm i}_{\beta}  \; 
\hat{\overline{u}}^{\rm j}_{\gamma}  \; \hat{u}^{\rm j}_{\delta}  \, , 
\hat{\overline{\psi}}_p^{\alpha_1} ({\bf r}_1 , 0) \bigg]_{-} \Bigg]_{+} 
| 0 \rangle  
\nonumber\\
&& = \frac{1}{6} \; 
\langle \hat{\overline u} \hat{u} \rangle_0 \;  \Bigg(
3 \, (\gamma_0)_{\delta \alpha_1} \; (\gamma_0)_{\alpha_2 \gamma} \; 
\delta_{\alpha \beta}  
\nonumber\\
&& -  (\gamma_0)_{\delta \alpha_1} \; (\gamma_0)_{\alpha_2 \alpha} \; 
\delta_{\beta \gamma} 
+ 3 \, (\gamma_0)_{\beta \alpha_1} \; (\gamma_0)_{\alpha_2 \alpha} \;
\delta_{\gamma \delta}     
\nonumber\\
&& -  (\gamma_0)_{\beta \alpha_1} \; (\gamma_0)_{\alpha_2 \gamma} \;
\delta_{\alpha \delta} \Bigg)\;,
\label{appendixC_5}
\end{eqnarray}
where the normalization (\ref{normalization_2}) and 
$\langle \hat{\overline u}^{{\rm i}}_{\alpha} \hat{u}^{\rm j}_{\beta}\rangle_0
= \frac{1}{12} \delta^{{\rm i}\,{\rm j}} \, \delta_{\alpha \beta} 
\;\langle \hat{\overline u} \hat{u} \rangle_0$ \cite{sumrule} 
has been used. 
For four-quark condensates with Gell-Mann matrices involved we find   
\begin{widetext}
\begin{eqnarray}
&&\int d^3 {\bf r}_1\;{\rm e}^{i {\bf k}_1 \, {\bf r}_1} \; \int d^3 {\bf r}_2\;
{\rm e}^{- i {\bf k}_2 \, {\bf r}_2}\;
\langle 0| \Bigg[ \hat{\psi}_p^{\alpha_2} ({\bf r}_2 , 0) \,,\bigg[
\hat{\overline{u}}^{\rm i}_{\alpha}  \; \hat{u}^{\rm j}_{\beta}  \;
\hat{\overline{u}}^{\rm k}_{\gamma}  \; \hat{u}^{\rm l}_{\delta}  \, ,
\hat{\overline{\psi}}_p^{\alpha_1} ({\bf r}_1 , 0) \bigg]_{-} \Bigg]_{+}
| 0 \rangle \; \left(\lambda^a\right)^{{\rm i}\, {\rm j}} \; 
\left(\lambda^a\right)^{{\rm k}\, {\rm l}} 
\nonumber\\
&& = |A_p|^2 \;
\epsilon^{{\rm a}\,{\rm b}\,{\rm c}} \epsilon^{{\rm a'}\,{\rm b'}\,{\rm c'}} \;
\left(u^{{\rm a'}\; {\rm T}} \, C \gamma_5 \, d^{\rm b'} \right) 
\left({\overline u}^{{\rm a}\; {\rm T}} 
\, C \gamma_5 \, {\overline d}^{\rm b} \right)
\;\left(2 \, \delta^{{\rm i}\, {\rm l}} \; \delta^{{\rm k}\, {\rm j}} - 
\frac{2}{3} \, \delta^{{\rm i}\, {\rm j}} \; 
\delta^{{\rm k}\, {\rm l}} \right) 
\nonumber\\
&&\times \; \Bigg(
(\gamma_0)_{\delta \alpha_1} \; (\gamma_0)_{\alpha_2 \gamma} \;
\delta_{\alpha \beta} \; \delta^{{\rm c} \, {\rm l}} 
\delta^{{\rm c'}\,{\rm k}} \delta^{{\rm i} \, {\rm j}} \;
- (\gamma_0)_{\delta \alpha_1} \; (\gamma_0)_{\alpha_2 \alpha} \;
\delta_{\beta \gamma} \; \delta^{{\rm c} \, {\rm l}} 
\delta^{{\rm i} \, {\rm c'}} \delta^{{\rm j} \, {\rm k}} \;
\nonumber\\
&&+ (\gamma_0)_{\beta \alpha_1} \; (\gamma_0)_{\alpha_2 \alpha} \;
\delta_{\gamma \delta} \; \delta^{{\rm j} \, {\rm c}} 
\delta^{{\rm i} \, {\rm c'}} \delta^{{\rm k} \, {\rm l}} \;
- (\gamma_0)_{\beta \alpha_1} \; (\gamma_0)_{\alpha_2 \gamma} \;
\delta_{\alpha \delta} \; \delta^{{\rm j} \, {\rm c}} 
\delta^{{\rm c'} \, {\rm k}} \delta^{{\rm i} \, {\rm l}}
\Bigg) 
\nonumber\\
&& = - \frac{8}{9} \; \bigg[ (\gamma_0)_{\delta \alpha_1} \; 
(\gamma_0)_{\alpha_2 \alpha} \; \delta_{\beta \gamma} 
+ (\gamma_0)_{\beta \alpha_1} \; (\gamma_0)_{\alpha_2 \gamma} \;
\delta_{\alpha \delta}\bigg] \; \langle \hat{\overline u} \hat{u} \rangle_0.
\label{appendixC_6}
\end{eqnarray}
\end{widetext}
By using the same technique the general result for the six-quark condensates 
is obtained as 
\begin{widetext}
\begin{eqnarray}
&&\int d^3 {\bf r}_1\;{\rm e}^{i {\bf k}_1 \, {\bf r}_1} \; \int d^3 {\bf r}_2\;
{\rm e}^{- i {\bf k}_2 \, {\bf r}_2}\;
\langle 0| \Bigg[ \hat{\psi}_p^{\alpha_2} ({\bf r}_2 , 0) \,,\bigg[
\hat{\overline{u}}^{\rm i}_{\alpha}  \; \hat{u}^{\rm j}_{\beta}  \;
\hat{\overline{u}}^{\rm k}_{\gamma}  \; \hat{u}^{\rm l}_{\delta}  \;
\hat{\overline{u}}^{\rm m}_{\epsilon}  \; \hat{u}^{\rm n}_{\zeta}  \; , \;
\hat{\overline{\psi}}_p^{\alpha_1} ({\bf r}_1 , 0) \bigg]_{-} \Bigg]_{+}
| 0 \rangle
\nonumber\\
&& = |A_p|^2 \;  
\epsilon^{{\rm a} \, {\rm b}\, {\rm c}}\; 
\epsilon^{{\rm a'}\, {\rm b'} \, {\rm c'}} \;
\left(u^{{\rm a'}\;{\rm T}} \, C \gamma_5 \, d^{\rm b'} \right)
\left({\overline u}^{{\rm a}\; {\rm T}} 
\, C \gamma_5 \, {\overline d}^{\rm b} \right)\;
\Bigg( \langle \hat{u}^{\rm j}_{\beta} \; \hat{\overline u}^{\rm k}_{\gamma} \;
\hat{u}^{\rm l}_{\delta} \; \hat{\overline u}^{\rm m}_{\epsilon} \; \rangle_0 \;
(\gamma_0)_{\alpha_1 \zeta} \; (\gamma_0)_{\alpha_2 \alpha}
\; \delta^{{\rm c}\, {\rm n}} \, \delta^{{\rm c'}\, {\rm i}}
\nonumber\\
&&+ \langle \hat{\overline u}^{\rm i}_{\alpha} \; \hat{u}^{\rm j}_{\beta} \;
\hat{u}^{\rm l}_{\delta} \; \hat{\overline u}^{\rm m}_{\epsilon} \rangle_0
\;(\gamma_0)_{\alpha_1 \zeta} \; (\gamma_0)_{\alpha_2 \gamma}
\, \delta^{{\rm c}\, {\rm n}} \, \delta^{{\rm c'}\, {\rm k}}
+ \langle \hat{\overline u}^{\rm i}_{\alpha} \; \hat{u}^{\rm j}_{\beta} \;
\hat{\overline u}^{\rm k}_{\gamma} \;  \hat{u}^{\rm l}_{\delta} \rangle_0
\;(\gamma_0)_{\alpha_1 \zeta} \; (\gamma_0)_{\alpha_2 \epsilon}
\; \delta^{{\rm c}\, {\rm n}} \delta^{{\rm c'}\, {\rm m}}  
\nonumber\\
&&+ \; \langle \hat{u}^{\rm j}_{\beta} \; \hat{\overline u}^{\rm k}_{\gamma} \;
\hat{\overline u}^{\rm m}_{\epsilon} \; \hat{u}^{\rm n}_{\zeta} \rangle_0
\;(\gamma_0)_{\alpha_1 \delta} \; (\gamma_0)_{\alpha_2 \alpha}
\, \delta^{{\rm c}\, {\rm l}} \, \delta^{{\rm c'}\, {\rm i}}
+ \langle \hat{\overline u}^{\rm i}_{\alpha} \; \hat{u}^{\rm j}_{\beta} \;
\hat{\overline u}^{\rm m}_{\epsilon} \;  \hat{u}^{\rm n}_{\zeta} \rangle_0
\;(\gamma_0)_{\alpha_1 \delta} \; (\gamma_0)_{\alpha_2 \gamma}
\; \delta^{{\rm c}\, {\rm l}} \delta^{{\rm c'}\, {\rm k}}
\nonumber\\
&&- \; \langle \hat{\overline u}^{\rm i}_{\alpha} \; \hat{u}^{\rm j}_{\beta} \;
\hat{\overline u}^{\rm k}_{\gamma} \; \hat{u}^{\rm n}_{\zeta} \rangle_0
\;(\gamma_0)_{\alpha_1 \delta} \; (\gamma_0)_{\alpha_2 \epsilon}
\, \delta^{{\rm c}\, {\rm l}} \, \delta^{{\rm c'}\, {\rm m}}
+ \langle \hat{\overline u}^{\rm k}_{\gamma} \; \hat{u}^{\rm l}_{\delta} \;
\hat{\overline u}^{\rm m}_{\epsilon} \;  \hat{u}^{\rm n}_{\zeta} \rangle_0
\;(\gamma_0)_{\alpha_1 \beta} \; (\gamma_0)_{\alpha_2 \alpha}
\; \delta^{{\rm c}\, {\rm j}} \delta^{{\rm c'}\, {\rm i}}
\nonumber\\
&&- \; \langle \hat{\overline u}^{\rm i}_{\alpha} \; \hat{u}^{\rm l}_{\delta} \;
\hat{\overline u}^{\rm m}_{\epsilon} \; \hat{u}^{\rm n}_{\zeta} \rangle_0
\;(\gamma_0)_{\alpha_1 \beta} \; (\gamma_0)_{\alpha_2 \gamma}
\, \delta^{{\rm c}\, {\rm j}} \, \delta^{{\rm c'}\, {\rm k}}
- \langle \hat{\overline u}^{\rm i}_{\alpha} \; 
\hat{\overline u}^{\rm k}_{\gamma} \;
\hat{u}^{\rm l}_{\delta} \;  \hat{u}^{\rm n}_{\zeta} \rangle_0
\;(\gamma_0)_{\alpha_1 \beta} \; (\gamma_0)_{\alpha_2 \epsilon}
\; \delta^{{\rm c}\, {\rm j}} \delta^{{\rm c'}\, {\rm m}}\Bigg)\;.
\label{six_10}
\end{eqnarray}
\end{widetext}
Note that for applying the normalizations (\ref{normalization_2}) 
and (\ref{normalization_3}) one needs a term $\delta^{{\rm c}\, {\rm c'}}$, 
which naturally 
arises when evaluating a specific matrix element under consideration.    


\begin{thebibliography}{99}
\bibitem{pcac3} The PCAC reads $\partial^{\mu} \hat{A}_{\mu}^a = 
f_{\pi} \, m_{\pi}^2 \, \hat{\varphi}^a$, where $\hat{A}_{\mu}^a$ is the 
axial vector current and $\hat{\varphi}^a$ is the pion field $a$.
\bibitem{pcbc6} The PCBC reads $\partial^{\mu} \hat{B}_{\mu}^{\alpha} = 
v \, \hat{\Psi}^{\alpha}$, where $\hat{B}_{\mu}^{\alpha}$ is the 
baryon current, $v$ is a proportional factor, and $\hat{\Psi}^{\alpha}$ 
is a baryon field with Dirac index $\alpha$. The PCBC can be generalized 
to baryon multipletts, so that $v$ becomes a matrix \cite{pcbc7}.
\bibitem{footnote1} Actually, in every special case under consideration one has
carefully to examine 
the vanishing of the nondiagonal terms. Otherwise one has to
to take into account them.
\bibitem{footnote2} Here, main contribution means that it gives about 50 percent 
of the total result. This is actually the case for most of the quark operators
we are going to consider, but in general not for operators which also contain
gluon fields like the nucleon mass operator.
\bibitem{footnote3} When replacing the nucleon field operator $\hat{\Psi}_N$
by a composite field operator $\hat{\psi}_N$ with quark degrees of freedom
one has to keep in mind that the equations of motion for both of these
field operators, in general, differ from each other, i.e. the space-time
evolution for these operators is different. However, in our investigation we
consider nucleon matrix elements of a local operator at the origin of space-time,
$\hat{\cal O} (x=0)$, where, according to Eq.~(\ref{eq_40}) and according to
the anticommutator relations for nucleon and quark fields $\delta$-functions,
$\delta^{(4)}(x_1) \, \delta^{(4)}(x_2)$, 
are implied both for $\hat{\Psi}_N$ and
$\hat{\psi}_N$, i.e. the space-time evolution for both field operators
becomes irrelevant.
\bibitem{numberoperator}  
The particle density operator $\hat{n}$ is defined as zeroth component of the 
current operator $\hat{j}_{\mu} (x) = \hat{\overline q} (x) \gamma_{\mu} 
\hat{q} (x)$. 
For a pointlike (bare) nucleon the volume $V \rightarrow 0$, so that 
$\hat{n} \rightarrow \hat{q}^{\dagger} (0) \, \hat{q} (0)$.
\bibitem{footnote4} When coupling two quarks to a diquark one obtains
$\{3\}_c \otimes \{3\}_c = \{6\}_c \oplus \{\overline{3}\}_c$, i.e. a totally
symmetric color sextet $\{6\}_c$ and a totally antisymmetric color antitriplet
$\{\overline{3}\}_c$. To get a color singlet nucleon state only the antisymmetric
color states $\{\overline{3}\}_c$, which are also color attractive,
cf. \cite{superconductivity},  can contribute. Therefore we have
$u^{{\rm a}\;{\rm T}} \, C \gamma_5 \, d^{\rm b} = -
u^{{\rm b}\;{\rm T}} \, C \gamma_5 \, d^{\rm a}$.
Then, together with the relation
${\overline u}^{{\rm a}\;{\rm T}} \, C \gamma_5 \,
{\overline d}^{{\rm b}} = \left( u^{{\rm a}\;{\rm T}} \, C \gamma_5 \,
d^{\rm b}\right)^{*}$
one may easily show that the term (\ref{normalization_2}) is a positive
real number. In the same way one proves that (\ref{normalization_3}) is also
a positive real number. Note that $A_p$ and $B_p$ have (mass dimension)$^{-3}$
and are adjusted such that the normalizations (\ref{normalization_2}) 
and (\ref{normalization_3}) are obtained.
\bibitem{footnote6} From neutron $\beta$ decay experiment, the total axial charge 
is $g_A =1.255$. A numerical value for the valence quark contribution
of the axial charge can be obtained by means of the decomposition
$g_A = g_A^v + g_A^{\pi}$ and the results of the MIT Bag model
(for massless pions), $g_A^{\pi} = 0.5 \, g_A^{v}$ \cite{hosaka},
which yields $g_A^{v} \simeq 0.84$.
\bibitem{footnote8} Taking into account the first two terms in
(\ref{factorization_5}) refers to the so called factorization approximation.
These terms, however, vanish when 
$\hat{A}$ and $\hat{B}$ are nonscalar operators.
On the other side, the decomposition (\ref{factorization_5}) is, by virtue
of a Fierz transformation of $\hat{A} \hat{B}$, not unique 
(see also footnote \cite{footnote7}).
Then, in particular, the factorization approximation does not vanish, even
in the case when $\hat{A}$ and $\hat{B}$ are nonscalar operators,
cf. Eqs.~(\ref{four_5}) and (\ref{four_10}).
\bibitem{footnote7} As an explicit example of footnote \cite{footnote8} 
we consider the vector channel, 
$\hat{\cal O}_V = \hat{\overline q}^{\rm i} \gamma_{\mu} \hat{q}^{\rm i} \; 
\hat{\overline q}^{\rm j} \gamma^{\mu} \hat{q}^{\rm j}$. 
Due to $\langle 0| \hat{\overline q}^{\rm i} \gamma_{\mu} \hat{q}^{\rm i} 
|0 \rangle = 0$ 
the factorization term vanishes according to Eq.~(\ref{factorization_5}), 
i.e. only the correction term remains: 
$1) \; \langle N_{\rm phys} | \hat{\cal O}_V | N_{\rm phys} \rangle = 
0 \,+ \,{\rm correction}\,(a)$. 
On the other side, a Fierz rearrangement of $\hat{\cal O}_V$ and 
afterwards the use of (\ref{factorization_5}) leads to 
$2) \; \langle N_{\rm phys} | \hat{\cal O}_V | N_{\rm phys} \rangle =  
- 2 \langle 0| \hat{\overline q}^{\rm i} \, \hat{q}^{\rm j} |0 \rangle\; 
\langle N_{\rm phys} | \hat{\overline q}^{\rm j} \, \hat{q}^{\rm i} 
| N_{\rm phys} \rangle 
+ \;{\rm correction}\,(b)$, 
i.e. the factorization term is not vanishing anymore. 
The seeming difference concerning the factorization term 
between the expression $1)$ and $2)$  
is absorbed into the correction terms $(a)$ and $(b)$  
which differ from each other. 
Furthermore, while the expression 1) is in line with 
Eq.~(\ref{factorization_5}), the 
expression 2) just coincides with Eq.~(\ref{four_5}).  
This concrete example elucidates the fact, that (\ref{factorization_5}) 
on one side and (\ref{four_5}), (\ref{four_10}) on the other side 
are matched with each other.  
Note that $\langle 0 | \hat{\overline q}^{\rm i} \hat{q}^{\rm j} | 0 \rangle 
= 1/3\;\delta^{{\rm i}\, {\rm j}} \; 
\langle 0 | \hat{\overline q} \hat{q} | 0 \rangle$, 
and $\langle N_{\rm phys} | \hat{\overline q}^{\rm i} \hat{q}^{\rm j} 
| N_{\rm phys} 
\rangle
= 1/3\;\delta^{{\rm i} \,{\rm j}} \; 
\langle N_{\rm phys} | \hat{\overline q} \hat{q} | 
N_{\rm phys} \rangle$.
\bibitem{footnote9} The results
of Ref. \cite{nuclear_matter_2} allow to extract the nucleon matrix elements
by means of the dilute nucleon gas approximation, where
$\langle\hat{\cal O}\rangle_{n} = \langle \hat{\cal O}\rangle_0 \, + \,
\langle N_{\rm phys} | \hat{\cal O} | N_{\rm phys} \rangle \, n/(2 M_N)$, 
and $n$ is the nucleon density.
\bibitem{footnote11} We compare the sign and magnitude of our evaluated valence
quark contribution with the total contribution for the physical nucleon,
because in  Ref. \cite{matrixlement_nucleon_5}
the valence quark contribution vanishes for all 
nonscalar four-quark condensates.
This vanishing is neither in contradiction with our findings nor with the
valence quark contribution extracted from the factorization approximation
in Ref. \cite{nuclear_matter_2} (see also \cite{footnote8, footnote7}).
\end{thebibliography}
\end{document}